%% file: main.tex
\documentclass[
	article,			
	12pt,				
	oneside,			
	a4paper,			
	english,			
	]{abntex2}

\usepackage{lmodern}
\usepackage[T1]{fontenc}		
\usepackage[utf8]{inputenc}		
\usepackage{nomencl} 			
\usepackage{color}
\usepackage{graphicx}			
\usepackage{microtype} 			
\usepackage{float}				
\usepackage{xspace}
\usepackage[defaultmathsizes,italic]{mathastext}
\usepackage{algorithmic}
\usepackage{algorithm2e}
\usepackage{amsthm}
\usepackage{amsmath}
\usepackage{amsfonts}
\usepackage{caption}
\usepackage{subcaption}
\def\lc{\left\lfloor}   
\def\rc{\right\rfloor}
\usepackage{lipsum}				
		
\date{}
\title{Dynamic MTU : Technique to reduce packet drops in IPv6 network resulted due to smaller path mtu size}
\autor{Ishfaq Hussain\textsuperscript{†}, \and Dr. Janibul Bashir\thanks{Corresponding author. Email: janibbashir@nitsri.ac.in }\thanks{Department of Information Technology, National Institute of Technology, Srinagar, India}}

\definecolor{blue}{RGB}{41,5,195}

\makeatletter
\hypersetup{
		colorlinks=true,       		
    	linkcolor=blue,          	
    	citecolor=blue,        		
    	filecolor=magenta,      		
		urlcolor=blue,
		bookmarksdepth=4
}
\makeatother

\makeindex

\setlrmarginsandblock{2cm}{2cm}{*}
\setulmarginsandblock{2cm}{2cm}{*}
\checkandfixthelayout

\SingleSpacing

\makepagestyle{meuestilo}
  \makeoddhead{meuestilo} 
     {\thepage}
     {}
     {HUSSAIN \textsc{et al.}}

\begin{document}

\textual

\pagestyle{meuestilo}

\frenchspacing 

\maketitle

\thispagestyle{meuestilo}

\begin{changemargin}{1cm}{1cm} 
 \textbf{Abstract} – With an increase in the number of internet users and the need to secure
internet traffic, the unreliable IPv4 protocol has been replaced by a more
secure protocol, called IPv6 for Internet system. The IPv6 protocol does not allow intermediate
routers to fragment the on-going IPv6 packet. Moreover, due to IP tunneling,
some extra headers are added to the IPv6 packet, exceeding the packet size
higher than the maximum transmission unit (MTU), resulting in increase in packet drops.
 One probable solution is to find the MTU of every link in advance
using the \textit{Internet Control Message Protocol} (ICMP) packets and accordingly fragment the packets at the source
itself. However, most of the intermediate routers and the network firewalls do
not allow ICMP packets to traverse through their network, resulting in network
black holes, where we cannot know the MTU of some links in advance. This method
tries to handle the packet drops in IPv6 network by proposing a DMTU scheme
where we dynamically adjust the MTU of each link depending upon the original
size of the IPv6 packet, thereby reducing the number of packet drops by a
significant amount. Using mathematical and graphical analysis, our scheme
proves to be much more efficient than the state-of-the-art PMTUD scheme. In this paper the method, mathematical and graphical representations are focusing solely in IPv6 Internet communication.

 \vspace{\onelineskip}
 
\end{changemargin}


\input{introduction}
\input{relatedwork}
\input{implementation}
\input{math}
\input{graphical}
\input{issues}
\input{conclusion}




\section*{References}

\vspace{-8mm}

\bibliographystyle{unsrt}
\bibliography{main}

\textbf{Dr. Janibul Bashir} has received his Ph.D in Computer Science and Engineering from Indian Institute of  Technology Delhi in 2019. He is currently working as an Assistant Professor in the Information Technology Department at National Institute of Technology, Srinagar. Before joining NIT Srinagar, he was working as a Software Engineer at Samsung Research India, Bangalore where he was awarded with Spot award at Samsung for developing 10 applications for the Tizen platform (Smartphone platform). He was also awarded with Gold Medal from NIT Srinagar in 2014. He current research is improving the performance of multicore systems, on-chip security, and application of machine learning techniques in the computer architecture domain (emerging technologies, network-on-chip, thermal management). He has also extended research interest in the operating systems and parallel programming (distributed systems). He can be reached at janibbashir@nitsri.ac.in.

\textbf{Ishfaq Hussain} was previously a student of National Institute of Technology, Srinagar with field of study concentric to Information Technology. He has a broad area of research interests in computer science and mathematical areas but not concentric to computer network, algorithms, mathematical analysis and computer architectures. He is currently a part of research community by contributing high authentic work to research communities of computer and mathematical society. He can be reached at ishfaqhussain90@gmail.com

\end{document}

%% file: introduction.tex
\section{Introduction}

Due to rapid increase in the number of devices connected to the internet, the current addressing scheme \textit{Internet Protocol Version 4} (IPv4) won’t be able to provide addresses to all the devices \cite{rfc4632}. It is because IPv4 uses 32-bit addresses, thus, less than 5 billion devices can be addressed using this scheme. And this limit has already been crossed by the number of internet enabled devices. Thus, there is a need of a new addressing scheme. Fortunately, \textit{Internet Corporation for Assigned Names and Numbers} (ICANN) has already announced a replacement to the IPv4 protocol, called IPv6 \cite{rfc8200}. It uses 128-bit addressing scheme, which can address upto $2^{128}\ (\approx 340\ undecillion)$ end devices \cite{rfc8200}. Besides, addressing trillions of devices, IPv6 provides a frame- work for a secure communication over an insecure network. It does this by using an add-on security protocol, called IPSec \cite{rfc4891}. For the proper functioning of IPSec, the IPv6 does not allow intermediate routers to fragment the on-going IPv6 packets. This security function is missing in the IPv4 protocol making it vulnerable to certain attacks. These security issue and addressing problem in IPv4 makes the new IP protocol v6 deployment very essential to keep on the smooth working of the Internet system.
 
Current migration from IPv4 to IPv6 is done through the transition, using principle of tunneling and translation techniques \cite{tunnel} which provides a means to carry IPv6 packets over unmodified IPv4 routing infrastructures and only a very small percentage of the network holds native IPv6. The process of tunnel came at a cost of increased packet size which pushes packet size greater than IPv6 minimum link \textit{maximum transmission unit} (MTU) \cite{rfc4291}. Moreover, a unique feature of NO fragmentation of packet with minimum link MTU of 1280 octets in IPv6 packet which further facilitates the intermediate nodes to drop the packet, which are having low path MTU in their outgoing links. 
 
There are many factors of packet drop by intermediate nodes namely network congestion , Bit Error Rate , TCP MSS size, low Bandwidth, collision. The \textit{Internet Control Message Protocol version 6} (ICMPv6) is  designed to inform the source nodes of such causes of packet drop by effecting nodes. Based on ICMP messages source node follows different mechanisms in sending the new regenerated packets. Namely, Congestion-control, Hamming-code, FEC, ALOHA and CSMA/CD ~\cite{IEEE} which addresses to the specific problem of packet drop in the link and many related new researches are proposed to improve these strategies in references \cite{conjestion,errorR,collision}. Similarly, the packet drop due to path MTU also make use of ICMP error packet, the intermediate effecting node sends ICMP PTB message to source node to re-transmit the packet. On bases of ICMP PTB message different solution can be used by source node.
 
One possible solution is to find the MTU of every possible link, using the ICMP
packets, in advance and accordingly fragment the packets at the source itself
~\cite{rfc8200} . However, most of the intermediate routers and the network
firewalls do not allow ICMP packets to traverse through their network,
resulting in network black holes, where we cannot know the MTU of some links in
advance. This scheme reduces the frequency of packet drops, but due to the
presence of black holes, the number of packet drops is still higher than the
tolerable limit.

The other solution is the \textit{Path MTU Discovery} (PMTUD) protocol~\cite{rfc8201}. To
avoid the IP fragmentation \cite{frag}, PMTUD determines the relevant MTU between the two
IP hosts and accordingly fragments the packet at the source itself. However,
this scheme relies on the ICMP packets, which has various issues related to the
black holes \cite{BB12} (please see Section~\ref{sec:related}).

To overcome the issues, we proposed a technique called Dynamic MTU (DMTU) where
we dynamically adjust the MTU of each link depending upon the original size of
the IP packet. After receiving the packet, we dynamically adjust the MTU of the
outgoing links and then accordingly forward the packet. However, in some
scenarios we cannot adjust the MTU of a link above a certain value. In those
cases we discard the packet and inform the source to further fragment the
packet.
\\


\vspace{5em}
\paragraph*{Our main contributions:}
\begin{enumerate}[label=\Alph*]
\item We proposed a technique called DMTU protocol to reduce the frequency of packet drops. 

\item We proposed different versions of the protocol in order to increase the efficiency of the system and thereby reduce the contention incise the network.

\item We presented the mathematical evaluation of our scheme and analyzed the frequency of packet drops, time delay, throughput, and latency enhancements.

\end{enumerate}

The rest of the treatise is as follows. Section~\ref{sec:related} discusses the
related work and motivations. In Section  ~\ref{sec:design} described the working of our proposed DMTU scheme followed by Section~\ref{sec:case} discusses analysis on time delay of our proposed method with previous start-of-art method Path MTU Discovery. Graphical analysis and comparison results are
presented in Section~\ref{sec:graphical}, and the paper concludes in
Section~\ref{sec:conc} with conclusions and future work.

%% file: relatedwork.tex
\section{Related Work and Motivations}
\label{sec:related}
Since to deal with packet drop as a result of Path MTU for IPv6, a proposal was made by ~\cite{rfc8201} called "Path MTU Discovery version 6" which is upgraded version of state-of-arts proposed by \cite{rfc1191} for IPv4. This method works in side by side with fragmentation at source. The main idea of Path MTU Discovery is that the source node initially assumes that the effective MTU, called \textit{Path MTU} (PMTU), of a path is the (known) MTU of the first hop in the path. If any of the packets sent on that path are too large to be forwarded by some node along the path, that node will discard them and respond back with an ICMPv6 ”Packet Too Big” packet.

Upon receipt of such a message, the source node reduces its assumed Path MTU for the path and makes it equivalent to the MTU of the constricting hop as reported in the ”Packet too Big” message. Thus, the source fragments the packets to make the size equal to new path MTU. The process ends when the source node’s estimate of the PMTU is less than or equal to the actual PMTU \cite{rfc8201}.

This method is currently deployment for IPv6 network and is up and running, but it suffers from very complicated issues about which we will explain briefly some of major problems in later subsections.

Further work present in \cite{PMTU-ISIS} by Vijay Kumar  which is extended in \cite{PMTU-ISISE} by Z. Hu  et al. in IETF working group, where the node calculates the Path MTU of the link using the IS-IS routing protocol, by doing so the overhead  incurred in the existing path maximum transferable unit discovery mechanism is reduced and the same solution can be extended to other link state routing protocols as well \cite{PMTU-ISIS}. This proposal is still an internet draft and is not approved for a RFC by IETF and hence under development and experimental.

In 2007 \cite{rfc4821} have came up with a new approach of finding Path MTU released on IETF an internet working group under RFC 4821. This method is named as Packetization Layer Path MTU Discovery (PLPMTUD) which is an extension to RFC 1191 and RFC 1981 which are ICMP based Path MTU Discovery. The PLPMTUD works above the packetizaion layer in the transport layer (TCP, SCTP, RTP etc.) to dynamically discover the MTU of the path by probing with consecutively larger packets with each successful packet delivery give rise to the packet size equal to the probing packet, without depending on ICMP PTB packets which resolves many of robustness problems which are in the classical PMTUD defined in RFC 1191 and RFC 1981. The Implementation of PLPMTUD is complicated as the probing packet don't tell about if it is lost weather its due to the congestion or path MTU issues. Its only deployed in the Linux kernel from version 2.6.17 which is off by default \cite{PSC}.

Currently, in Sep 2020 an updated version of PLPMTUD is released named as Datagram Packetization Layer Path MTU Discovery (DPLPMTUD) in \cite{rfc8899}, which has same working as in PLPMTUD but works in UDP layer protocol.

Both of the New approaches PLPMTUD and DPLPMTUD need to send probing packets to know the path MTU of the link and is a hit and trail method of knowing the path MTU. Sending probing packets might help in successfully transmission of the original packet with minimum risk of loosing the packet, but with the price of consumption of network resources of sending multiple probing packets. These two methods still dependent on the source node and the usage of probing packets creates same issues like the ICMP-PTB packets which contributes to increased time delay for transmitting a packet. Unlike Classical PMTUD, PLPMTUD and DPLMTUD, our proposed method work on intermediate node level to solve the path mtu with no need of ICMP-PTB or probing packet at source level. The proposed method is designed in such a way that it will work in parallel mode with classical PMTUD  and in Standalone mode, without use of Classical PMTUD.

In this study PMTUD is refereed to classical PMTUD throughout the paper which is defined in RFC 1191 and RFC 1981 and don't implies to the PLPMTUD or DPLPMTUD. Again the PMTUD mentioned in the parallel mode algorithm is referring to the classical PMTUD and motivations for designing the proposed method is also from the failures seen in PMTUD which is currently the working mechanism and didn't take in-account the PLPMTUD and DPLPMTUD which still needs to resolve and review many issues for implementation. 

In the following subsections we have briefly elaborated some of major issue of classical PMTUD which motivated the need of developing a new proposed mechanism which works at intermediate node level in standalone and in parallel with classical PMTUD.

\subsection{Path MTU Discovery Vulnerability}

 The Path MTU Discovery is vulnerable to two types of denial of service attack (DOS). A DOS attacks on a networking structure to restrain or stop the servers from communicating to its clients \cite{dos,dosaic}. Both of these attacks are based on malicious party sending false \textit{Packet Too Big} (PTB) message by informing smaller or larger Path MTU size than in reality to the source node \cite{rfc8201}. In each of these attacks the sender observes sub-optimal performance and temporary blockage and can cause Black hole connection, where TCP hand shaking performs completely but connection hangs during data transfer ~\cite{rfc8201}.
 A single iteration of PMTUD procedure doesn’t completely assure the transmission of packet to the destination it needs to go for multiple iteration of PMTUD algorithm to successfully sends the packet to destination. Hence the frequent use of the PMTUD by the source node in sending the same packet results in increase in the fragmented packs of small size and consumption of network resources.
 
 \subsection{Network Resources Consumption}
 
 The nodes using Path MTU Discovery must need to detect decreases in PMTU as sooner as possible. To keep a Path MTU updated the source node continues to elicit Packet Too Big messages then the current estimated path MTU after a time frame of 5 minutes or 10 minutes (as Recommended), this updating can likely be prone to stale and false packets that were floating in the network and those source nodes having multiple paths to the destination, where each carries different Path MTU might result in updating false Path MTU ~\cite{rfc8201}. Beside this, the updating process in which source continue to elicit messages leads to consumes network resources ~\cite{rfc8201}.
 
 In \textit{Path MTU Discovery} (PMTUD), packets which are very small in size are forwarded through the network for a long period of time which may result in inefficient usage of the network bandwidth which is the common case in the PMTUD scheme, when path has a node which has much lower MTU then the other nodes  ~\cite{rfc8201, PMTU-ISIS}.
 
 The PMTUD uses ICMP message to know the Path MTU of the link , but the ICMP packet has to travel from problem occurred node to the source node which consumes considerable amount of bandwidth on all the intermediate links between the problem occurred node and the source node \cite{PMTU-ISIS}.
 
\subsection{Path MTU Black Holes}

In~\cite{BB12} claims that the main cause for the occurrence of PMTUD black holes in the Internet is the filtering of important signalling packets by the intermediate nodes on the path. In the event that these nodes run a firewall, they could potentially be configured to disallow all or certain types of Internet Control Message Protocol (ICMP) packets to pass through them. The effect of this filtering results in the inoperable of \textit{Path MTU Discovery} (PMTUD) and this failure of PMTUD will lead to black holes \cite{PMTUDUNI}.
 
 Most of the routers give low priority to the ICMP messages carrying the information of the dropped packets. In addition, most of the users and ISP providers configure their firewalls to block all ICMP messages~\cite{PMTUDUNI} because it wastes the network resources and creates congestion in the network \cite{rfc1435}. Thus, due to low priority and blocking of ICMP messages, the ICMP messages sent in the PMTUD will be either delayed or blocked from reaching the source node, resulting in either the increased delay or the ICMP black-holes \cite{rfc2923,PMTUDUNI}.
 
 This will lead to inoperable and connectivity failure. Due to ICMP message unreachable problem, the host will continuously keep on sending packets and will eventually lead to increased congestion in the channel, thereby reducing the overall performance of the system. 
 
 \subsection{Path MTU Discovery Incompatibility with IPv6}

The Path MTU Discovery \cite{rfc1191}, with the base design is made keeping the features and characteristics of the IPv4 protocol without taking IPv6 characteristics in account. Suppose, if a link has Path MTU lesser then minimum Path MTU of IPv6 packet then in that case how the situation can be handled? In such scenario, sending nodes violate the IPv6 constrains by generating packets of size lower than the prescribed minimum Path MTU of IPv6 packet \cite{rfc8200} widely in IPv6 network tunnelling, which in further results in wastage of bandwidth usage ~\cite{rfc4459}. Which make Path MTU Discovery far more less effective in handling packet drop due to Path MTU in IPv6 network. Keeping this issue in one hand another factor that slows the PMTUD in IPv6, that the ration of lowest path MTU of IPv4 to the standard MTU of link is much smaller than the ratio of lowest path MTU of IPv6 to the standard MTU of link. i.e.

\[\left({LowestPathMTU(IPv4 ) \over standard(MTU)} << {LowestPathMTU(IPv6) \over standard( MTU)}\right)\]

 Since we haven’t seen any change in Standard MTU value for IPv6 network , after the minimum packet size raised from 576 octets to 1280 octets in IPv6 \cite{rfc8200}. In prior IP version even a lower packet size of 576 bytes can be forwarded or transmitted \cite{rfc791} without any fragmentation or packet drop, while that’s not in case of IPv6. Like a case in DNS which don’t accepts packet size greater than 512 bytes ~\cite{rfc7719} and a IPv6 source node is not allowed to create or generate a packet of size lower than 1280 octets which makes DNS incompatibility issue in IPv6 network \cite{dns}. This shows that a small range of packet size from 1280 to 1500 octet can only be transmitted to the IPv6 network. While packets greater than 1500 octet gets an ICMP error message and packets lower than 1280 octets are restricted \cite{rfc8200}. Since, there is no any other strategy for the source node other then fragmentation of the packet after getting ICMP PTB message due to Path MTU and in some cases, it is not allowed to do fragmentation of packets which are not a factor of 1280 octet because it results in generating packet of size lower than 1280 octet \cite{rfc1981}. In IPv4 network, if a source node generates a packet in range of 576 to 1500 octets and during transmission the packet encountered a node having smaller MTU then it has the permission to fragment the packet if Don’t fragment bit is not been set by the source node \cite{rfc791}. Thus PMTUD in IPv4 has to handle only packets which have DF bit set which is not in case of IPv6 network where packet can’t be fragmented by intermediate nodes \& the source node can’t fragment lower than 1280 octet which further makes the IPv6 packets vulnerable to be dropped by the nodes in the path and make increase in workload of PMTUDv6 then in PMTUDv4. Despite such compatibility issues, PMTUDv6 in IPv6 network operates in a large range of packet size $Y$ such that, \[Y : U - \{1280,\ldots, 1500\}\] \[where, \qquad U \in \mathbf{N}^+\] whereas, in PMTUDv4 in IPv4 network operates in a smaller range of packet size $X$ respective to PMTUDv6 such that, \[X : U - \{576,\ldots, 1500\}\]
 In general which further makes Path MTU Discovery less effective and low efficient technique in IPv6 network.

 Since from all of these drawback in previous proposed inventions (\cite{rfc8201},  \cite{rfc1981} and \cite{PMTU-ISIS}) no new invention or scheme is initiated or proposed in intermediate node level to get through these issues in IPv6 protocol. In this invention we initiate an algorithm in the intermediate node on receiving such large packets then Path MTU on the fly to prevent packet drop and further optimise the load due to re-transmissions, probing, maintenance of Path MTU value and fragmentation's in source node. In our proposed method we tried to reduce the usage of ICMP message to a far very low or even to null value. In this method, the intermediate nodes process the packet on the fly without violating any network policies and regulations, and hence reaches to maximum throughput and lower latency of network. Thereby rescuing a lots of packets in IPv6 network to be the victim of Path MTU.

%% file: implementation.tex
\pdfoutput=1
\section{Proposed Mechanism}
\label{sec:design}
 In this paper we present a Dynamic MTU scheme, in which it dynamically adjust the MTU of the intermediate links based on the size of the received packet. As soon as the intermediate router transmits a packet to the next forwarding interface link, it reverts back the MTU to its original value. In Figure~\ref{process} describes the DMTU algorithm in two phases: Pre-DMTU and Post-DMTU phases. The algorithm starts with the Pre-DMTU phase and then moves to Post-DMTU. In the Pre-DMTU phase, the algorithm checks the basic conditions necessary for the DMTU functionality such as whether the MTU of the intermediate links can be changed or not. When the Pre-DMTU signals the success (all conditions are satisfied) then only the algorithm moves to the Post-DMTU phase, otherwise it stops the algorithm and drops the packet and informs the receiver about the same. The main aim behind dividing the algorithm into two phases is to avoid the unnecessary computations in case any necessary condition is violated. Thus, decreasing the overheads on the node. 
 
 \begin{figure}[!b]
\centering
\includegraphics[width=0.7\columnwidth]{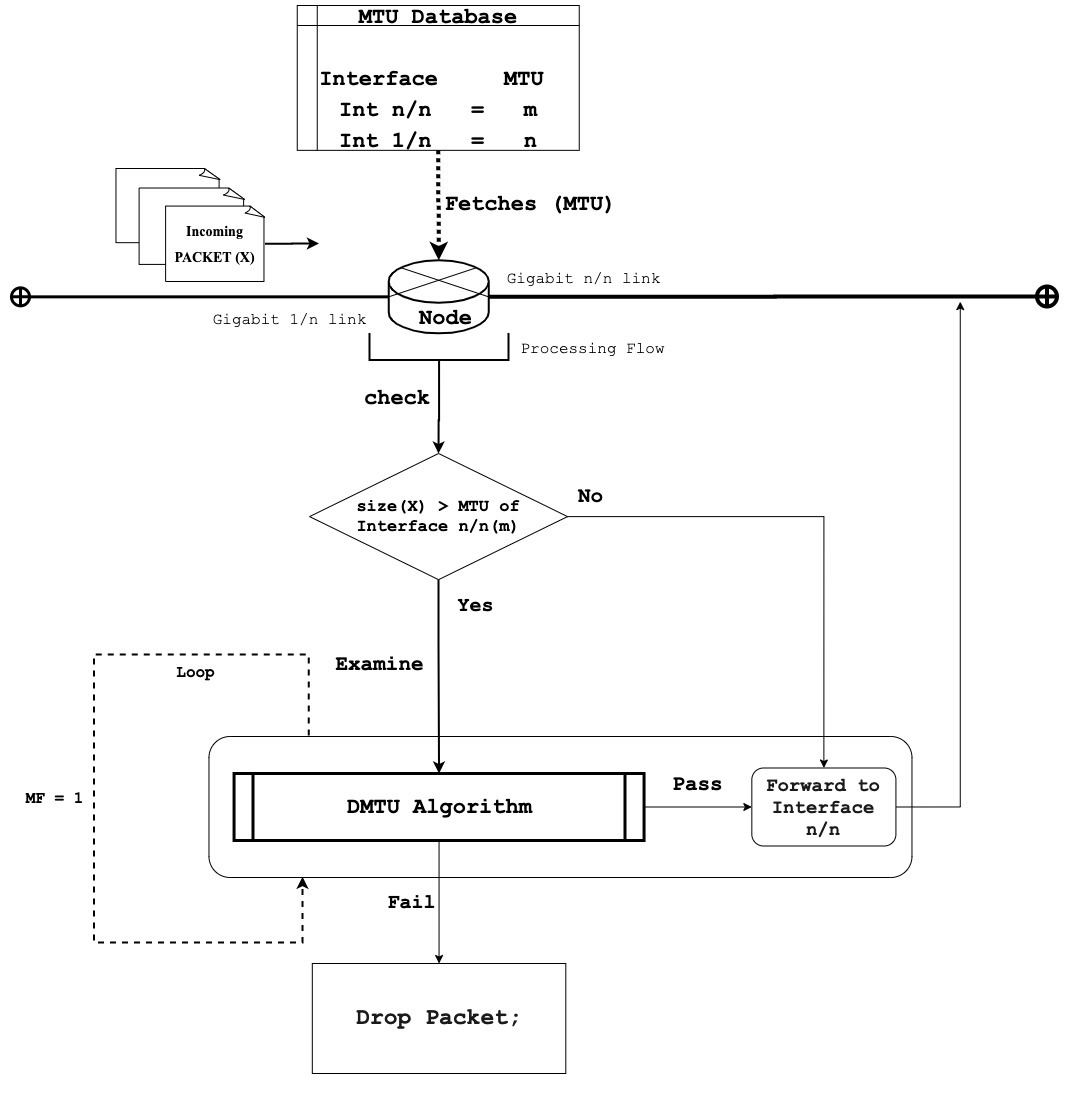}
\caption{DMTU Mechanism inside the node}
\label{layer}
\end{figure}

 In Figure~\ref{layer} illustrates the working of DMTU algorithm inside an intermediate node in general form, that when it receives a packet larger than then next forwarding link MTU, it examine the packet then based on the examination it either passes the packet to next Interface link n/n or drops the packet on failing the examination. If a flow has multiple packets then the algorithm is called recursively using More fragment bit property of packet only, to overcome the load and time in examining.
 
 The DMTU algorithm can be used either in parallel with the PMTUD scheme or it can be used as a standalone scheme. In the former case, we call the scheme as Pre-Parallel DMTU  and the later one is called Standalone DMTU. 
In Pre-Parallel DMTU algorithm, we allow DMTU algorithm to run only for high priority packets and whereas all other packets are handled using the PMTUD scheme. However, in the standalone-DMTU, DMTU completely replace the PMTUD algorithm and takes the charge.

Note that the there is a minor difference between the Pre-Parallel DMTU and Standalone DMTU algorithms. The main motive behind having two schemes is to allow the algorithm to work in the networks where the PMTUD algorithm has already been deployed. 
 
 \subsection{DMTU Phases}
 The DMTU algorithm can be used either in parallel with the PMTUD scheme or it can be used as a standalone scheme. In the former case, we call the scheme as Pre-Parallel DMTU and the later one is called Standalone DMTU. In Pre-Parallel DMTU algorithm, we allow DMTU algorithm to run only for high priority packets and whereas all other packets are handled using the PMTUD scheme. However, in the standalone-DMTU, DMTU completely replace the PMTUD algorithm and takes the charge. Note that the there is a minor difference between the Pre-Parallel DMTU and Standalone DMTU algorithms. The main motive behind having two schemes is to allow the algorithm to work in the networks where the PMTUD algorithm has already been deployed.
\subsubsection{Pre-DMTU Phases}

\begin{figure}[!t]
\centering
\includegraphics[width=0.8\columnwidth]{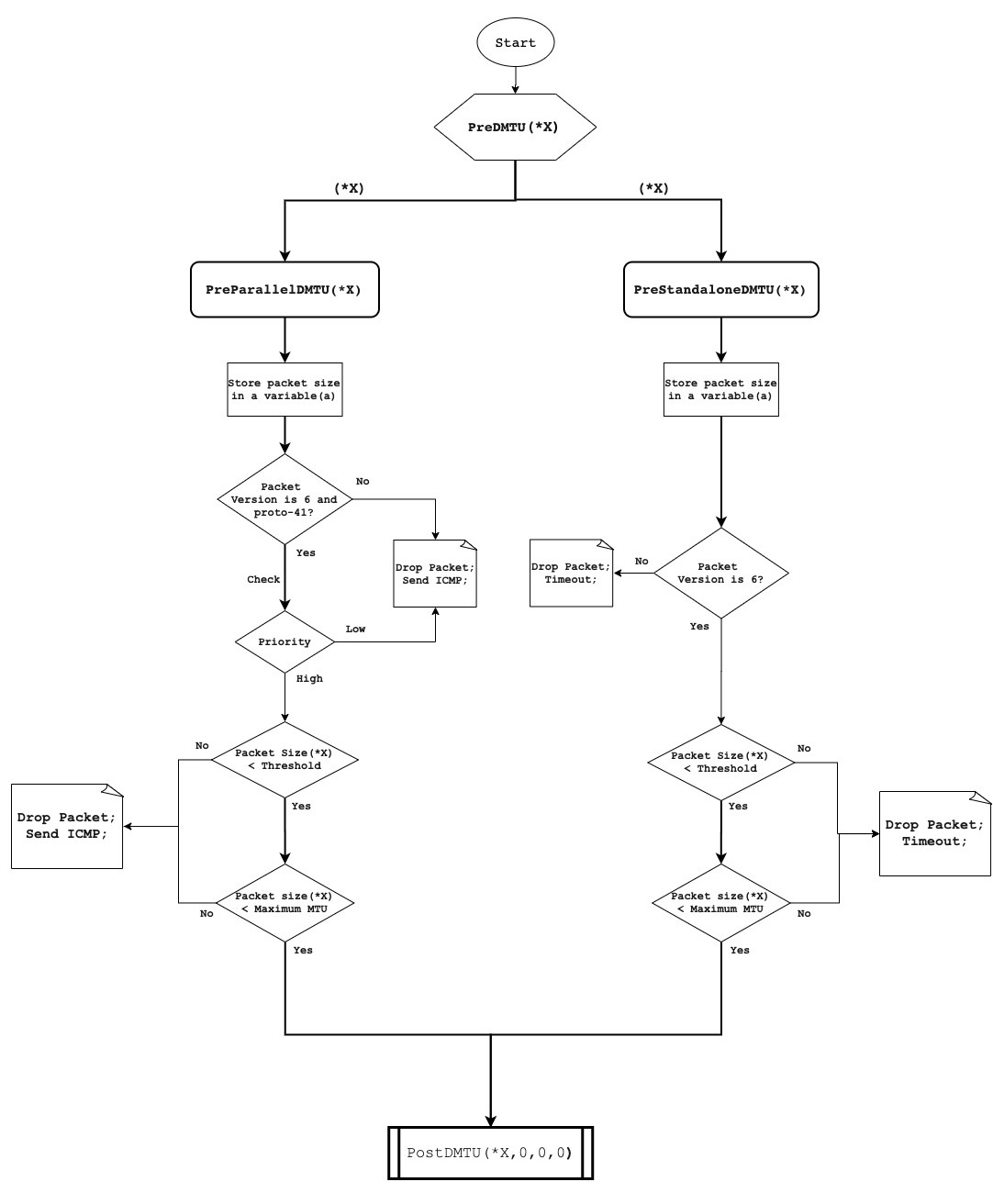}
\caption{Flow chart of Pre-DMTU Phases.}
\label{predmtu}
\end{figure}

 Figure~\ref{predmtu} illustrates the working of Pre-DMTU Phases. This phases carries one argument $(^*X)$ as the pointer to the packet in the buffer memory. This phase triggers on receiving an incoming packet of size greater than the next-interfaces MTU. There are two type of phases, PreStandalone DMTU phase and PreParallel DMTU phase, both of which works differently depending upon the type of strategy we are assuming. Both of the phases carries a pointer to the packet $(^*X)$ from the buffer memory to be processed. These phases act as an preliminary examination of the packet to run the PostDMTU phase.

\textbf{Pre-Standalone DMTU Phase:}
 Algorithm~\ref{Algo-Standalone} shows a pseudo-code of the PreStandalone DMTU phase: It works in parallel with Path MTU Discovery. When a new packet invokes, it compares the size of the packet with the threshold and Maximum MTU limit of the path. If the packet size exceeds these limits then it is not possible to successfully route the packet. Instead we force the router to drop the packet (without informing the source). The source will itself re-transmit the packet by fragmenting it into smaller packets after the timeout. However, if the packet size is lesser than the Maximum-MTU then the algorithm moves to the Post-DMTU phase.
 
   \begin{algorithm}
\DontPrintSemicolon
\caption{PreStandaloneDMTU(*X)}
\textbf{Input: } \text{A pointer to the packet $X$ at buffer memory such that $Size(^*X) > MTU_{n/n}^{Curr}$}\\
\textbf{Output: }  \text{Parameters $\alpha$, $\beta$\ and\ $C$\ with\ a\ pointer\ $^*X$ to PostDMTU}\\

     $a$ $\xleftarrow{\text{store}}$ $Size(^*X);$\\
      \eIf{$Ver = 6$}{
      \eIf{$a \le threshold\ \&\&\ a\ \le\   MTU_{n/n}^{Max}$}{
      $PostDMTU(^*X,0,0,0);$
      }{
      $TIMEOUT;$
      }
     }{
     $TIMEOUT;$
     }
\label{Algo-Standalone}
\end{algorithm}

\textbf{Pre-Parallel DMTU Phase:}
Algorithm ~\ref{Algo-Parallel} shows the pseudo-code of PreParallel DMTU phase which allows DMTU to work only for the high priority packets in IPv6 and Tunnelled packets with \textit{proto-41}. 

\begin{algorithm}
\DontPrintSemicolon
\caption{PreParallelDMTU(*X)}
\textbf{Input: } \text{A pointer to the packet $X$ at buffer memory such that $Size(^*X) > MTU_{n/n}^{Curr}$}\\
\textbf{Output: }  \text{Parameters $\alpha$, $\beta$\ and\ $C$\ with\ a\ pointer\ $^*X$ to PostDMTU}\\

     $a$ $\xleftarrow{\text{store}}$ $Size(^*X);$\\
      \eIf{$Ver = 6\ ||\ PN = 41$}{
      \eIf{$Priority = High$}{
      \eIf{$a \le threshold\ \&\&\ a\ \le\   MTU_{n/n}^{Max}$}{
      $PostDMTU(^*X,0,0,0);$}{
      $ICMP;$ }
      }{$ICMP;$}
     }{$ICMP;$}
\label{Algo-Parallel}
\end{algorithm}

Thus, the job of Pre-DMTU phase in this scheme is to first find the priority of the packets. Based on the packet priority it either directs the PMTUD algorithm to take the control or it moves to the Pre-DMTU next step.

 In Pre-DMTU next step, it compares the size of the packet with the threshold value. If the packet size is less than the threshold value then it moves to the $PostDMTU(^*X,C,\alpha,\beta)$ phase by returning parameters $(^*X,0,0,0)$, else it discards the packet and informs the PMTUD scheme to send an ICMP error message ”packet too big” to the source.
\\
\subsubsection{Post-DMTU Phase}

\begin{figure}[!t]
\centering
\includegraphics[width=0.8\columnwidth]{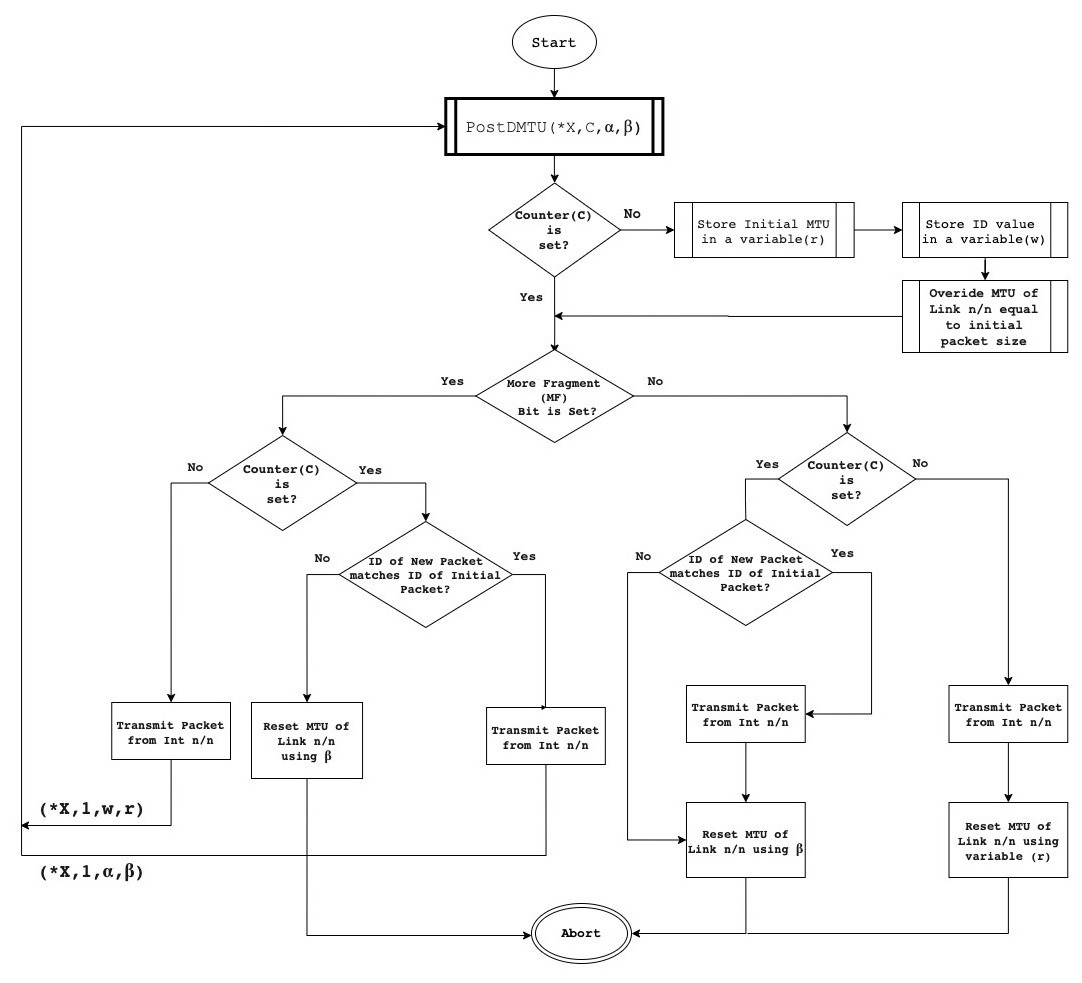}
\caption{Flow Chart of working of PostDMTU Phase}
\label{postdmtu}
\end{figure}

 Figure~\ref{postdmtu} \& algorithm~\ref{algopost} illustrates the working of PostDMTU phase, which is the heart of the algorithm. Its main function is to change the state of the MTU of the forwarding port to a higher state according to the incoming packet size up-to a threshold limit and forward s the packet through the next-Interface link, after that reverting MTU of the forwarding link to its initial state.
 
 When a packet pass the PreDMTU phase, then it will be the invoking packet of PostDMTU algorithm, the algorithm then follows next steps by storing the initial/original MTU of next-Interface and Identification number of packet in some variables, followed by overriding the MTU of the forwarding-Link according to the packet size at location $(^*X)$ in buffer memory. Since, the invoking packet may be the fragment of a flow or a single packet without multiple fragments, therefore using the MF bit from the header of the packet, the algorithm follows different steps.

\begin{figure}[!t]
\centering
\includegraphics[width=0.5\columnwidth]{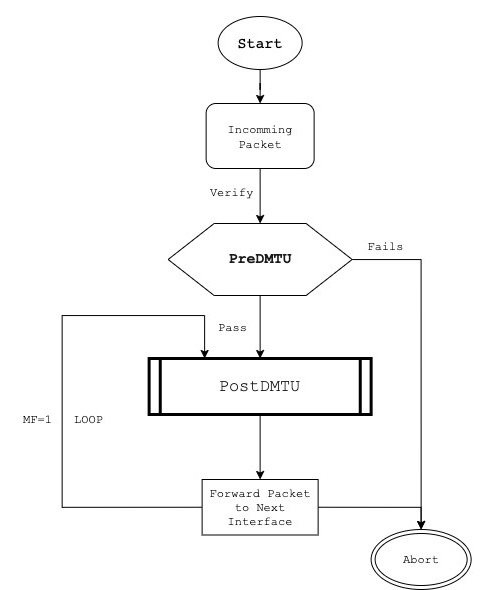}
\caption{Flow chart of DMTU for multiple fragments.}
\label{process}
\end{figure}

 If there are multiple fragmented packets from a flow than for every fragment we need to do an overriding of MTU followed by PreDMTU phase examination which could take a large network resource and processing time. That's why we take care of this multiple fragments of same flow using parameter triplets $(C,\alpha,\beta)$ also figure~\ref{process} describes the working of DMTU Phases with Multiple fragments.

\textbf{a). At More fragment = 0: }
 If the MF bit is not set i.e MF=0, then the packet may be the initial packet or the final packet of a flow. Both types of packet overrides the next interface MTU according to the packet size and then forwards the packet to next interface after which it resets the MTU to its initial state. But the packet which is last packet of a flow needs to compare its ID with initial packet of the same flow if fails it reset the MTU to initial state.

\textbf{b). At More fragment = 1: }
 If MF bit is set i.e $MF=1$, then it forwards the packet to next-interface link. After forwarding the packet in location $^*X$ in buffer memory the location is now empty and the next packet in the queue takes the location of the previous forwarded packet. To process multiple packets/fragments we makes a recursive call to itself, on doing so it again repeat the same process for the new arriving packet from the buffer memory of storing ID , MTU of next interface and overriding the next Interface MTU. But initial PostDMTU call have already changed the MTU of the next interface MTU, therefore the current running MTU of the interface is not the original MTU and no need to store it in any variable also its not feasible to override again and again the MTU of the next interface according to the packets of same size, its enough to override the MTU according to the initial invoking packet of the flow. Hence, to settle down all these issues we use a parameter C called counter whose value is in \{0 or 1\} in PostDMTU.
 
 \begin{algorithm}
\DontPrintSemicolon
\caption{PostDMTU($^*X,C,\alpha,\beta$)}
\textbf{Input: } \text{Parameters $\alpha$, $\beta$\ and\ $C \in \{0,1\}$. A Pointer to packet $X$ such that $Size(^*X) > MTU_{n/n}^{Curr}$}\\
\textbf{Output: }  \text{Forward packet at $^*X$ to $Int_{n/n}$}\\
\If{$C=0$}{
$Var\ r$ $\xleftarrow{\text{store}}$ $MTU_{n/n}^{Curr};$\\
$Var\ w$ $\xleftarrow{\text{store}}$ $ID(^*X);$\\
$MTU_{n/n}^{Curr}$ $\xleftarrow{\text{override}}$ $Size(^*X);$\\
}
      \eIf{$MF=1$}{
      \eIf{$C=1$}{
      \eIf{$ID(^*X)=\alpha$}{
      $PostDMTU(^*X,1,\alpha,\beta);$}{
     $MTU_{n/n}^{Curr}$ $\xleftarrow{\text{reset}}$ $\beta;$\\
     }
      }{
      $Int_{n/n}\ \xleftarrow{\text{forward}}$ $^*X$;\\
      $PostDMTU(^*X,1,w,r);$
      }
     }{
     \eIf{$C=1$}{
      \eIf{$ID(^*X)=\alpha$}{
      $Int_{n/n}$ $\xleftarrow{\text{forward}}$ $^*X;$\\
        $MTU_{n/n}^{Curr}$ $\xleftarrow{\text{reset}}$ $\beta;$\\
      }{
      $MTU_{n/n}^{Curr}$ $\xleftarrow{\text{reset}}$ $\beta;$\\
      }
     }{
     $Int_{n/n}$ $\xleftarrow{\text{forward}}$ $^*X;$\\
     $MTU_{n/n}^{Curr}$ $\xleftarrow{\text{reset}}$ $r;$\\
     }
     }
\label{algopost}
\end{algorithm}
 
 If Counter value is not set then it means it is the first call from PreDMTU phase, else it is recursive call by PostDMTU which would happen only if there are multiple packets. Therefore after forwarding the initial packet it returns position of next packet as $(^*X)$ and values of initial MTU and ID of the initial invoking packet with a counter value $C = 1$, identifying it as recursive call by PostDMTU. When a recursive call is made to PostDMTU then the next packet to process may be last fragment or may not. If it is not the last packet then this time the algorithm compares the ID of new arriving packet with initial packet from parameter $\alpha$. 
 
 The comparison of ID is done to keep running the postDMTU for the flow related to the initial Packet or if the packet router has multiple queues and uses Bit-by-Bit Round Robin (BBRR) then in that case it jumps from one queue to another then the ID of the new arriving packet may not be the same as the initial invoking packet and a PreDMTU examination is important before calling PostDMTU. if ID’s are not same then it reverts the MTU of port to its initial state using parameter $\beta$, which carries the original MTU of the port, else if ID’s are same then it forwards the packet to next interface link and then do a recursive call to PostDMTU. This time we didn’t stored the ID and MTU value in any variable and instead returning the variables we return the parameters carrying these values $alpha$ and $\beta$ to PostDMTU along with counter value set and location of next packet. Every recursive call carries the ID and the original MTU of port in $\alpha$ and $\beta$ and it ends when it encounters MF bit 0, which means this is last fragment of the flow and hence after forwarding this packet it reverts the port MTU to its initial state using parameter $\beta$.
 
 In brief, when the new packet arrives, it compares ID number of incoming packet with the stored ID value , if these are same then it forwards multiple packets until MF (more fragments) bit equals to zero. When MF=0, it means no more fragments of the packet are arriving. After this, the algorithm reverts back the MTU to the original value.
 
The working of Post-DMTU algorithm  is same for the Pre-Parallel DMTU and Pre-Standalone DMTU phase. 

%% file: math.tex
\pdfoutput=1
\section{Measuring the Time Delay}
\label{sec:case}

\begin{figure}[!t]
\centering
\includegraphics[width=1\columnwidth]{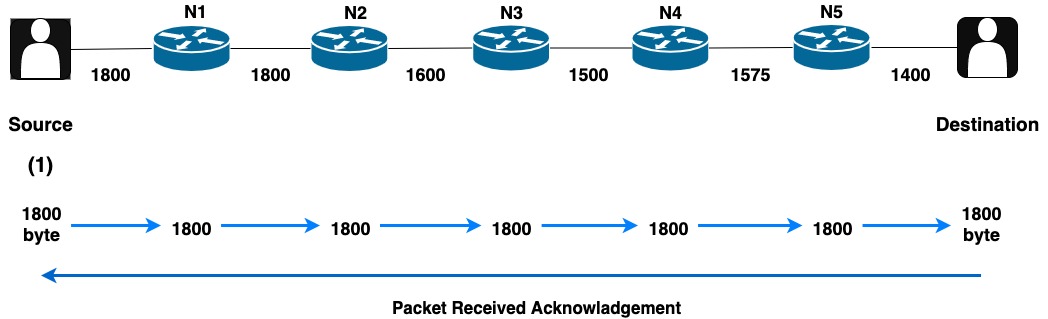}
\caption{ End to End Delay using DMTU.}
\label{DMTU_demo}
\end{figure}

In Internet system time is everything, its the basic unit to measure and compare the efficiency and Quality of a network, network parameters and network equipment. So we preferred to proceed first to measure the effects of proposed methods on time delay by comparing it with previous state-of-the-art PMTUD algorithm. Our analysis of time delay for DMTU algorithm make use of the study we carried-out in \cite{ishfaq} , which have taken a analysis on time delay in PMTUD algorithm in transmitting a single packet of 1800 bytes in a finite network configuration using PMTUD algorithm. We will take the same case study of sending 1800 bytes of packet in same network configuration but using Dynamic MTU algorithm rather Path MTU Discovery for the comparision results. In Figure~\ref{DMTU_demo} is a network configuration similar as taken in Figure~1 in \cite{ishfaq} for Path MTU Discovery, with same packet size for transmission of 1800 byte of packet from source to destination over a path of varying mtu size using Dynamic MTU algorithm.

The outputs from this case study will be used later in the paper, to further analyse other network parameters and to find out how better and efficient is the new proposed mechanism then the Path MTU Discovery. 

\subsection{Effect on Total Time Delay Using DMTU
}
As the analysis led by \cite{ishfaq} we will also begin by sending a packet of size 1800 bytes from source to destination over the same network configuration as shown in Figure~\ref{DMTU_demo}.
 The packet will passes through the first node N1 and reaches to node second (i.e N2), since the link mtu is lesser then the incoming packet N2 runs the DMTU algorithm and send the packet through the next interface of node N2 to node N3, where in before case study in \cite{ishfaq} the node N2 has dropped the packet and needs to send the packet again. Now the node N3 also runs the DMTU algorithm due to exceeding the packet size then its forwarding link mtu and transmits the same packet to next interface of node N3. Then the packet passes from node N4, and node N5 runs the DMTU algorithm and transmits the packet to destination. Since, on sending the packet from the source to destination the DMTU algorithm is triggered 3 times in the nodes where the next interface MTU is lesser the incoming packet size. Therefore, the time for running the algorithm by router is \(T_O\) and for three nodes it will be \(3T_O\)  time which would be added to the total transmission time. 

In DMTU case, the first transmission is the first successful transmission without the need of extra fragmentation of packet, while the similar case study in \cite{ishfaq} the $4^{th}$ transmission is the first successful transmission with packet size decreased by 400 bytes. 
This led to an increase of End-to-End-Delay (E2ED) between nodes \(^DT_D\) in DMTUD algorithm by a quantity of $^{D-P}T_D)$ from the E2ED of successful transmission using PMTUD algorithm in \cite{ishfaq}. Therefore, the E2ED of successful transmission $(i.e\ ^DT_D)$ in using DMTUD algorithm in terms of E2ED of successful transmission $(i.e\ ^PT_D)$ using PMTUD algorithm in \cite{ishfaq} is:

\begin{gather}
	^DT_D = (^PT_D + ^{D-P}T_D) \label{DMTU_delay}
\end{gather}

The second entity of the Equation ~\ref{DMTU_delay} i.e   $^{D-P}T_D$ is E2ED between nodes for the resultant packet size from subtracting packet size of successful transmission in PMTUD from DMTU i.e packet size in DMTU is taken as D and Packet size in PMTUD is taken as P , therefore the resultant packet size is superscript as \((D-P)\) and its E2ED is taken as \(^{D-P}T_D\), which is nearly equals to the difference in time delay of \(^DT_D\ and\ ^PT_D\). 

In play $^DT_D \ge \ ^PT_D$, therefore the \((^DT_D\ -\ ^PT_D)\) is positive and is $< \epsilon$.

\textit{i.e}
\begin{align}
  ^{D-P}T_D =| ^DT_D - ^PT_D| < \epsilon \label{lessepsil} \\ 
\implies \qquad ^{D-P}T_D < \epsilon \label{epsil}
\end{align}


The range of $\epsilon$ is not definite, throughout the paper we assume that it is a very small value in fractions that when $\pm \epsilon$ to any natural number $\mathbb{N}$ results in minor change after decimal values and the product with any natural number $\mathbb{N}$ results value approaches to approximately 0.

\textit{i.e}
\begin{align*}
 x\ \pm \epsilon\ \simeq\ x \\
x\ \epsilon\ \simeq\ 0\\
\forall\ x\ \in\ \mathbb{N}   
\end{align*}

The total time delay $^DT$ of the successful transmission depends upon the overhead time of the node $T_O$ and the E2ED $^DT_D$ and no fragmentation time as no fragmentation is carried-out using DMTU.
Therefore, the total time delay in DMTU algorithm would be:
\begin{gather}
	Total\ time\ delay\ (^DT) = (n + 1)^DT_D + 3T_O \label{Total_DMTU_}
    \end{gather}
    Since, there are only 5 nodes then the Equation \ref{Total_DMTU_} becomes:

\begin{gather}
^DT  = 6(^DT_D) + 3T_O
\end{gather}
From Equation \ref{DMTU_delay} the total time delay be:
\begin{align}
&^DT\ =\ 6(^PT_D + ^{D-P}T_D)+ 3T_O\notag\\
&^DT\ =\ 6(^PT_{D}) + 6(^{D-P}T_D) + 3T_O\\
where, \qquad 	&|^{D-P}T_D| < \epsilon \notag
\end{align}
Since, the total time wastage \(T_{W}\) will be
\begin{gather}
^DT_{W} = 6(^{D-P}T_D)+ 3T_O
\end{gather}
Therefore, total time delay can be represented as:
\begin{align}
	^DT &= 6(^PT_D)+ ^DT_{W}\notag\\
	^DT &= 6(^PT_D)+ 6(^{D-P}T_D)+ 3T_O \label{TimeDMTU} 
\end{align}
Since the study given by \cite{ishfaq} for analysis of PMTUD algorithm the corresponding time delay in PMTUD for the same network configuration is given by :
\begin{align}
    ^PT = (6(^PT_D) + 6(^{D-P}T_D) +3T_O) \label{pmtudcasepre}
\end{align}
Subtracting the Equation ~\ref{pmtudcasepre} from ~\ref{TimeDMTU} we get:
\begin{align}
^DT - ^PT & = (6(^PT_D) + 6(^{D-P}T_D) +3T_O) - (6(^PT_D) + 10(^PT_D +^PT’_D) + 3T_F)\\
^DT - ^PT & =(6(^{D-P}T_D) - 10(^PT_D)) + (3T_O  - (10T’_D  + 3T_F))
\end{align}
\textit{Since,}\\	
\[[6(^{D-P}T_D) - 10(^PT_D)]\   <\   0 [^{D-P}T_D < ^PT_D]\]
Also,
\[ [3T_O  - {10T’_D  + 3T_F}] < 0 \quad(\ :. [T’_D + T_F] > T_O)\]
Since, the outcome of the two negative value will be a negative value which implies that :
\[^DT_{W} < ^PT_{W} \]

Hence, the total time delay in DMTU is less than the total time delay in PMTUD in a same network configuration which implies that the use of DMTU decreases the time delay. 
From the  above calculations we can show that total time for n nodes dropping at a nodes can be given by:
\begin{gather}
Total\ Time = [n+1][^PT_D +^{D-P}T_D] + aT_O\\
where, \qquad 	|^{D-P}T_D| < \epsilon \notag
\end{gather}

\subsection{Analysis of Total Time Delay for $n$ - Nodes}

 The calculations provided in Section~\ref{sec:case} are limited to 3 nodes. However, 
the total time taken by the DMTU and PMTUD for n nodes between source to destination, considering packet drop at ‘a’ number of nodes at position $n_{i}th$ is given as:\\

For PMTUD as from findings of \cite{ishfaq}:
\begin{gather}
^PT = ^PT_D(n+1) + (^PT_D+^PT'_D)\sum_{i=1}^{\ a } [n_i]+aT_F \label{PMTUDs_Total}
\end{gather}

For DMTU:
\begin{gather}
^DT = ^PT_D(n+1)+ ^{D-P}T_D(n+1) + aT_O \label{DMTU_Total}
\end{gather}
Subtracting Equation \ref{PMTUDs_Total} from \ref{DMTU_Total} we have :

\begin{align}
^DT - ^PT &= \left (^{D-P}T_D(n+1) + aT_O \right )- \left ((^PT_D+^PT'_D)\sum_{i=1}^{\ a } [n_i] +aT_F\right )\notag\\
	&= \left (^{D-P}T_D(n+1) - (^PT_D+^PT'_D) \sum_{i=1}^{\ a } [n_i] \right ) + \left (aT_O- aT_F \right )\notag \\
^DT - ^PT	&= \left (^{D-P}T_D(n+1) - (^PT_D + ^PT’_D)\sum_{i=1}^{\ a } [n_i] \right )+ a \left (T_O - T_F \right ) \label{DMTU_PMTU}
\end{align}

Now the term \(\sum_{i=1}^{\ a } [n_i]\) creates two scenarios, consecutive and non-consecutive case scenarios. 
 In consecutive case scenario when packets drop in Consecutively i.e \(n_i - n_{i-1} = n_{i+1} - n_i\), where in non-consecutive the packet drop is random and is non-consecutive that is \(n_i - n_{i-1}\ne n_{i+1} - n_i.\)

\subsubsection{Non-Consecutive Case Scenario}

\begin{figure}[!t]
\centering
\includegraphics[width=0.7\columnwidth]{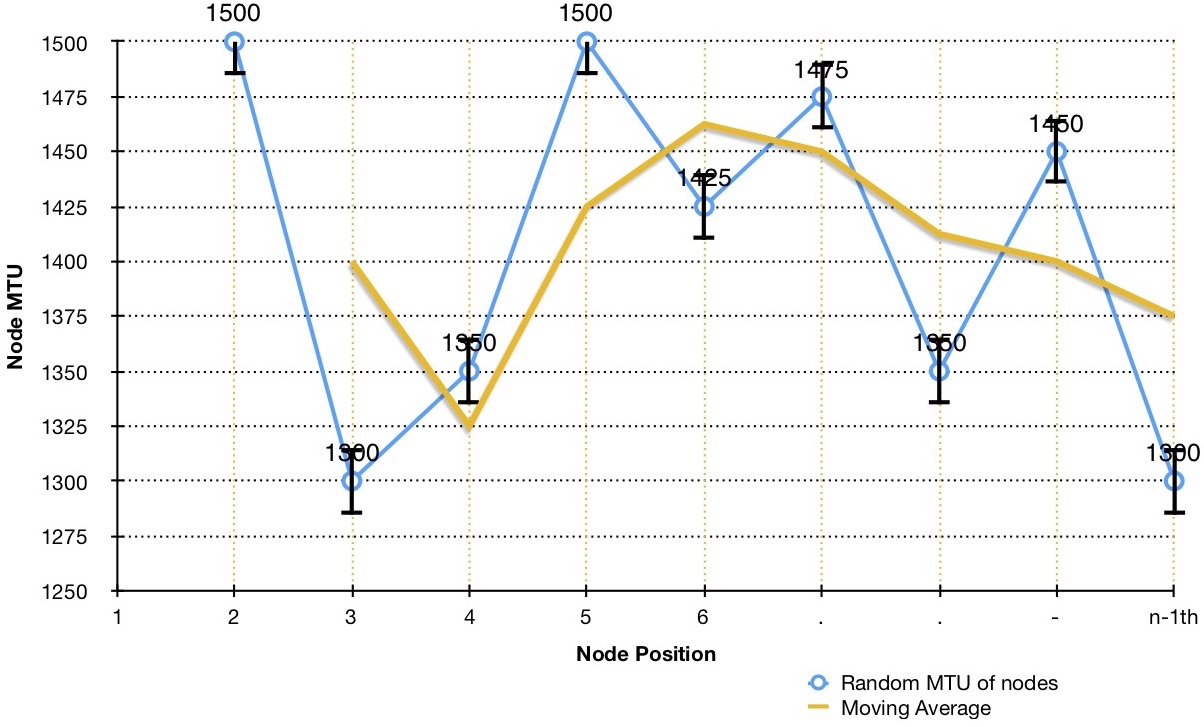}
\caption{ Representation of MTU's of the nodes in random fashion in non-consecutive case.}
\label{node:random}
\end{figure}

The non-Consecutive case scenario arises when the nodes drop packet randomly \(i.e\ n_i - n_{i-1} \ne n_{i+1} - n_i\) , then the term \(\sum_{i=1}^{\ a } [n_i]\) = \((n_1+n_2+n_3+n_4+…n_a)\). In Figure~\ref{node:random} is the representation of the MTU's of the nodes in random fashion in a network path in non-consecutive case.

Since, \(^{D-P}T_D << ^PT_D + ^PT’_D\) 
and the value of  \(^{D-P}T_D < \epsilon \).\\
Therefore, the effect of \((n+1)\ on\ |^{D-P}T_D|\) will be:
\begin{gather}
(n+1) ^{D - P}T_D < \epsilon\\
\implies \left ((n+1)^{D-P}T_D \right )  <  \left ((^PT_D + ^PT’_D)\sum_{i=1}^{\ a } [n_i] \right ) \label{D_P} 
\end{gather}
Hence, 	
\begin{gather}
\left ({^{D-P}T_D(n+1) - (^PT_D + ^PT’_D)\sum_{i=1}^{\ a } [n_i]} \right)  < 0
\end{gather}
Also,
\begin{gather}
\qquad T_O < T_F\notag \\
\implies  {T_O - T_F} < 0 \label{O_F}
\end{gather}

Since from Equations \ref{D_P} and \ref{O_F} the result of sum of two negative value is a negative therefore:
\begin{gather}
^DT - ^PT < 0\notag \\
\implies ^DT < ^PT	, \quad	\forall\ n \in Z^+ \label{D<P}
\end{gather}

\textit{Hence the total time in the DMTU \((^DT )\) is lesser then total time in PMTUD \((^PT)\) in random case scenario. }

\subsubsection{Consecutive Case Scenario}

\begin{figure}[!t]
\centering
\includegraphics[width=0.7\columnwidth]{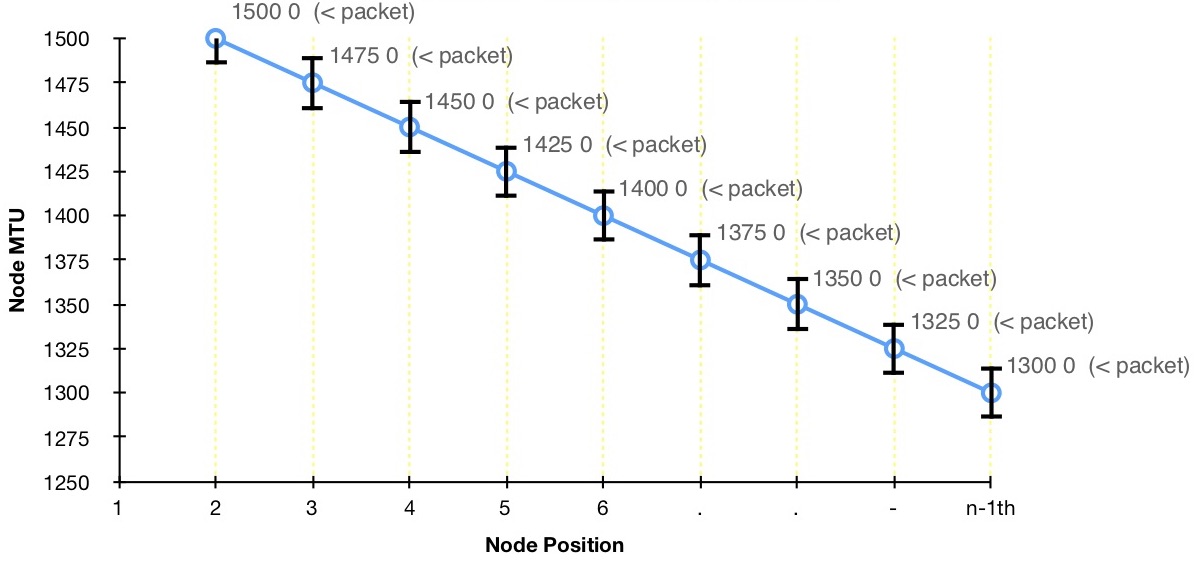}
\caption{ MTU's of nodes in series fashion in the network path in consecutive case.}
\label{node:series}
\end{figure}

In consecutive case scenario when nodes drop packet consecutively \(i.e\ n_i - n_{i-1} = n_{i+1} - n_i\)  then the term \(\sum_{i=1}^{\ a } [n_i] = S_a =  {a(a+1)\over2}\). In Figure~\ref{node:series} show the nodes with MTU in decreasing fashion which give rise to the consecutive packet drop by nodes.

Then the Equation \ref{DMTU_PMTU} will become \begin{align}
^DT - ^PT\ =\ ^{D-P}T_D(n+1) - (^PT_D + ^PT’_D)S_a + a(T_O + T_F) \label{Conse_D_P}
\end{align}
From Equation \ref{Conse_D_P} the term:\\ \[\left (^{D-P}T_D(n+1) - (^PT_D + ^PT’_D)S_a \right ) < 0,\]
\quad\(iff,\)
\begin{align}
&n+1 < {a(a+1)\over 2}\notag\\
\implies \qquad &2(n+1) < a^2 + a\notag\\
\implies \qquad &a^2 + a - 2n -2 < 0 \label{quad}
\end{align}
		
Since the Equation \ref{quad} becomes quadratic equation with roots \(\alpha > {-1+(9+8n)^{1\over2}\over2}\)  and \(\beta > {-1-(9+8n)^{1\over2}\over2}\) . From the two roots ,the root \(\beta > {-1-(9+8n)^{1\over2}\over2} <\)  0 which implies that\textit{ ‘a’ }has a negative value, which is not possible as \(a\ \subset\ n, n\in Z^+\). Therefore the root of quadratic equation is \(\alpha > {-1+(9+8n)^{1\over2}\over2} >\) 0. 
The root value has a fractional value and nodes can't be in fraction, therefore we take the floor value of root \(a\) i.e\\
\begin{align*}
    Floor(a)\ =\ \lc a \rc
  \end{align*}
Since the success probability of \(n+1 < {a(a+1)\over 2}\ at\ a > \lc {-1+{(9+8n)^{1\over2}}\over2} \rc\) is given by probability function:
\[P(n+1 >\sum_{i=1}^{\ a } [n_i])=1-{a-1\over n}\]
\(At\ a=  {-1+{(9+8n)^{1\over2}}\over2},\)
\[P(n+1 >\sum_{i=1}^{\ a } [n_i])
 \le 1-{[{-1+(9+8n)^{1\over2}\over2}] -1\over n}\]

\begin{figure}[!t]
\centering
\includegraphics[width=0.7\columnwidth]{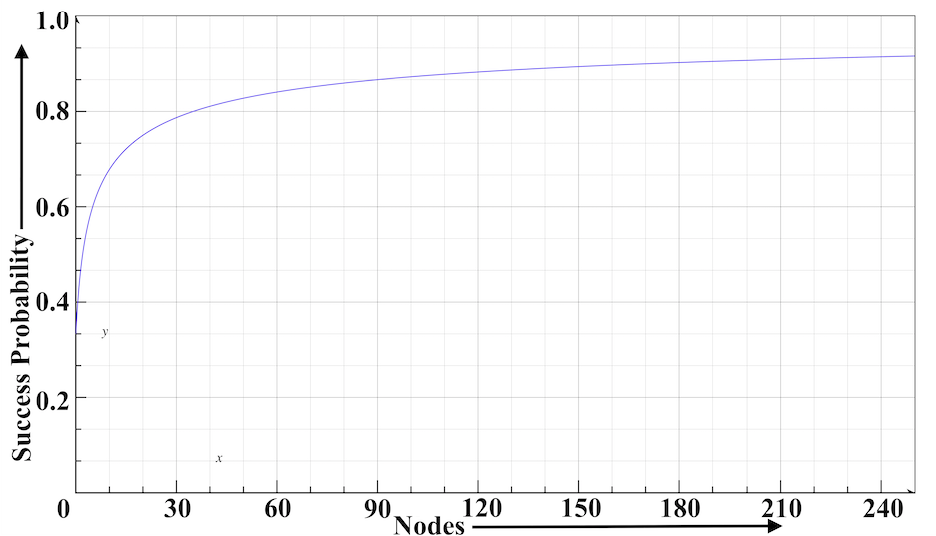}
\caption{ Success probability graph over number of nodes.}
\label{pro:sucess}
\end{figure}

In Figure~\ref{pro:sucess} shows the success probability by varying the number of nodes, which shows at highest success probability of \(log_e(2.5173)\) at nodes \(300\) and lowest success probability of \(log_e(1.39)\) at node 0, over the number of nodes the term \( {a(a+1)\over 2} > (n +1)\ at\ a\ > \lc {-1+{(9+8n)^{1\over2}}\over2} \rc > 0 ,\forall\ a \subset\ n\ and\ n\ \in\ Z^+\), which implies that: 
\[^{D-P}T_D(n+1) < (^PT_D + ^PT’_D) {a(a+1)\over 2}\].

\textit{Therefore,}
\[{^{D-P}T_D(n+1) - (^PT_D + ^PT’_D) {a(a+1)\over 2}} < 0. \]\\
Also, from Equation \ref{O_F} the term:

\begin{align}	
&{T_O - T_F} < 0 \notag \\
\implies\qquad &^DT < ^PT \label{D<P_Con}
\end{align}

In Table~\ref{data} the data is collected from the success probability graph over number of nodes  when \(n \ge n_i > \alpha ,\ where\  \alpha >  \lc {-1+(9+8n)^{1\over2}\over2} \rc\)\(  \in\ (n_i)\) at \(\sum_{i=1}^{\ a } [n_i]\) which shows success probability increases rapidly from 0.3333 to 0.99 as number of nodes increases for \((n+1) < {a(a+1)\over 2}\) 














\begin{center}
\begin{table}[h]%
\centering
\caption{Data from probability graph over nodes.\label{data}}%
\begin{tabular*}{500pt}{@{\extracolsep\fill}ccc@{\extracolsep\fill}}
\hline
\textbf{Number of Nodes} & \textbf{Success Probability} &  \textbf{In Terms of \(Log_e(s)\)}\\
\hline
0	&	0.3332	&	\(log_e(1.3954)\)\\
1	&	0.4384	&	\(log_e(1.5502)\)\\
5	&	0.6000	&	\(log_e(1.8221)\)\\
10	&	0.6783	&	\(log_e(1.9705)\)\\
20	&	0.7500	&	\(log_e(2.1170)\)\\
30	&	0.7870	&	\(log_e(2.1967)\)\\
40	&	0.8108	&	\(log_e(2.2497)\)\\
50	&	0.8278	&	\(log_e(2.2882)\)\\
100	&	0.8728	&	\(log_e(2.3936)\)\\
200	&	0.9072	&	\(log_e(2.4773)\)\\
300	&	0.9232	&	\(log_e(2.5173)\)\\
\hline
\end{tabular*}
\end{table}
\end{center}


\textit{Hence the total time in the DMTU \((^DT )\) is lesser then total time in PMTUD \((^PT)\) in consecutive case scenario with success probability of \(log_e(2.718)\) for n-nodes.}

%% file: graphical.tex
\pdfoutput=1
\section{Graphical Analyses and Comparison Results}
\label{sec:graphical}
To find out how much effective and robust is the new proposed method compared to the previous stat-of-art algorithm , we carried out a  graphical representation of the network parameters namely Time Delay, Latency and Throughput of both the methods by deriving the related equation of parameters and then compared them. 
To calculate and compare these parameters we need to find out how the general equations of total time delay of both the algorithms behaves, under some fixed pre-defined values and varying percentage of packet drop. For that we fix the values for the terms of both the general equations of total time delay to be the average/highest limit as seen in real network analyses software's, so that the results will be accurate.

In \cite{ishfaq} shows the complete measurements of the time-delay due to Path MTU discovery in IPv6 network, and we use those calculations for the comparison results. 
Since the parameters which we are fixing pre-defined value are given by;
\(^PT_D=10^{-1},\; ^{D-P}T_D=10^{-2},\; T_O=10^{-3},\; ^PT'_D=^PT_D-15\cdot 10^{-2},\;\)
By keeping these parameters constant and taking iterations of ‘a’ which is the nodes which dropped the packet i.e
\(a=\left(1 0\%n,\; 20\%n,\; 30\%n,\; 40\%n,\; 50\%n\right)\)

We will now be able to get a better graphical representation of these network parameters using these fixed pre-defined value and by varying percentage of packet drop.

\subsection{Graphical Analysis of Total Time Delay}
In Equation \ref{DMTU_Total} we have Total time delay for DMTU as: 
\begin{align}
^DT\; =\; n\left( ^PT_D+^{D-P}T_D \right) +\left(^PT_D+^{D-P}T_D+(aT_O)\right)\;\label{dmtu_grap}
\end{align}
The Equation \ref{dmtu_grap} is in form of \(y = ax + b,\) where \(a = (^PT_D+^{D-P}T_D)\) and \(b = (^PT_D+ ^{D-P}T_D + (aT_O))\)
Taking n towards x-axis and \(^DT\) towards y-axis which gives us a straight line, origin at intercept \((^PT_D+ ^{D-P}T_D + aT_O)\) of y-axis with (+) slope of \((^PT_D+^{D-P}T_D)\).

From Equation~\ref{PMTUDs_Total} we have Total time delay for PMTUD as:
\begin{gather}
^PT = ^PT_D(n+1) + (^PT_D+^PT'_D)\sum_{i=1}^{\ a } [n_i]+aT_F \label{PMTU_Total}
\end{gather}
 which also the form of  \(y = ax + b,\), with \(a = (^PT_D)\) and Intercept of \(b = ^PT_D + \sum_{i=1}^{\ a } [n_i]( ^PT_D+^PT'_D )+aT_F\), Where the value \(T_F \approx T_O\), therefore we take the value of \(T_F = T_O\) as we now don’t have the overhead time calculated but our assumption is that \(T_O < T_F\) and for the sake of calculation, we treat it to be equal to \(T_F\).
 Now both the equations follows the linear equation where number of nodes (n) are along x-axis and Total time delay \(^PT\) along y-axis. By subtracting \(^PT\) from \(^DT\) will also produce a linear equation as:

\begin{align}
^DT-^PT=\; n\left(^{D-P}T_D \right) + \left( \left(^{D-P}T_D \right) - \left( ^PT_D+^PT'_D \right)\left(\sum_{i=1}^{\ a } [n_i] \right) \right)\label{d-p}
\end{align}

Which has a intercept of \(\left( \left(^{D-P}T_D \right) - \left( ^PT_D+^PT'_D \right)\left(\sum_{i=1}^{\ a } [n_i] \right) \right)\) and \(a=n\left(^{D-P}T_D \right)\).

\begin{figure}[!t]
  \centering
  \begin{subfigure}{0.4\columnwidth}
    \includegraphics[width=1\columnwidth]{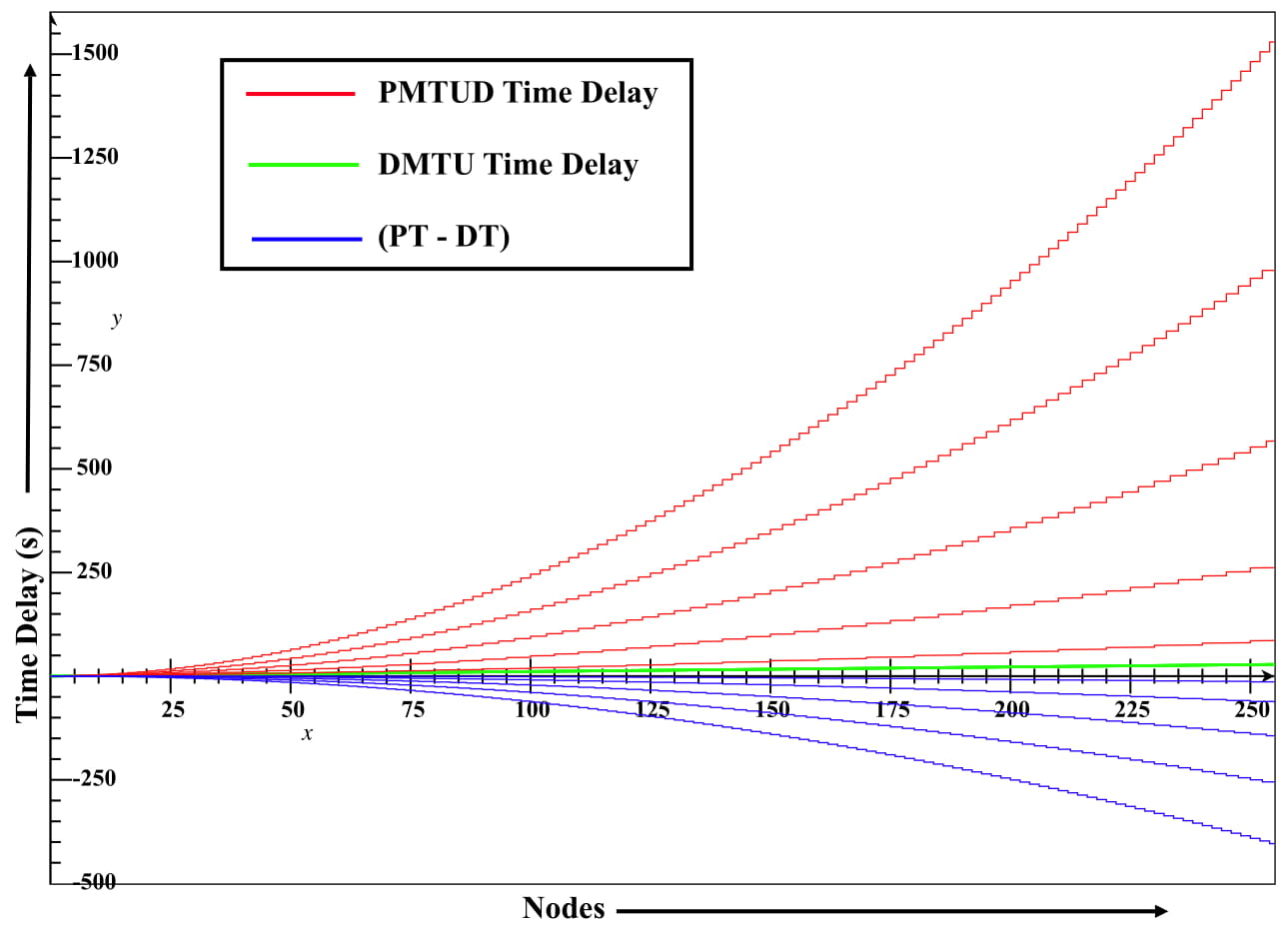}
    \caption{At Min(\(\sum_{i=1}^{\ a } [n_i]\))}
    \label{fig:delaymin}
  \end{subfigure}
  \begin{subfigure}{0.4\columnwidth}
    \includegraphics[width=1\columnwidth]{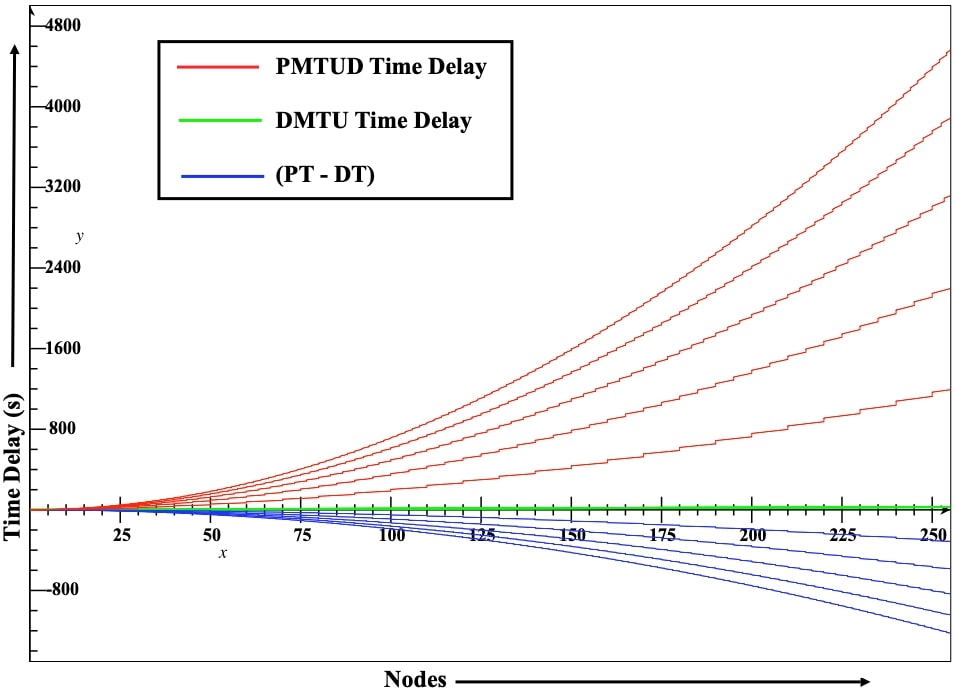}
    \caption{At Max(\(\sum_{i=1}^{\ a } [n_i]\))}
    \label{fig:delaymax}
  \end{subfigure}
  \caption{ Comparisons of time delay in DMTU and PMTUD.}
  \label{fig:coffee}
  \footnote{The graph shows that the straight line formed by Equation of DMTU is at lower altitude than straight line formed by equation of PMTUD, implies  DMTU has lower time delay than PMTUD ,also the difference of D-P is declined from 0 toward -ive y - direction for every value of x, hence \(^DT < ^PT\).}
\end{figure}

Now by applying the fixed pre-defined values in both the Equations~\ref{dmtu_grap}, \ref{PMTU_Total} \& \ref{d-p} by varying percentage of packet drop by nodes we get a spectrum of lines moving upwards, downwards and straight as shown in Figure~9a \& Figure~9b.

The Figure~\ref{fig:delaymin} is derived at minimum value of \(\sum_{i=1}^{\ a } [n_i]\) , which means the packet drop starts from first node and  consecutively till value of \textit{ 'a'} which is the last position of the node dropping the packet. In this analysis we see at minimum value of \(\sum_{i=1}^{\ a } [n_i]\) the time delay for the DMTU algorithm is lower than the PMTUD algorithm. The green line passing closer to the x-axis represents total time delay in DMTU algorithm, while the red lines going in upper directions shows the percentage of nodes dropping packets in different iterations in PMTUD algorithm and blue-line going downward direction along x-axis represents the amount of time delay saved by DMTU algorithm.

The Figure~\ref{fig:delaymax} is drawn at maximum value of \(\sum_{i=1}^{\ a } [n_i]\), i.e. at \(a = n\), which means the packet drops consecutively till the nth node in the path and starts from \textit{x-a} value, which represents the highest value of \(\sum_{i=1}^{\ a } [n_i]\) for nodes that drop packet. This analysis shows that the DMTU algorithm is effective for routes passing large number of nodes with greater probability of packet loss than using PMTUD.

\subsection{Graphical Analysis of Throughput Enhancement}

The DMTU algorithm increases the throughput with respect to the throughput without DMTU algorithm. And we will show the enhancement in the throughput mathematically.

\textit{Mathematically :}
\[Throughput={Data\ Transmitted \over Unit\ of\ Time}\]
Let \(^PT_R\) be throughput using PMTUD algorithm. Since,  Unit of Time = Total Time = \(^PT\) and
\( \  Data\ Transmitted\ =\ Packet\ Size\)\\
\[	\implies ^PT_R={Packet\ Size\over(^PT)}		\]
\textit{Let Packet size = P,}
\begin{gather}
\implies ^PT_R={P\over(^PT)} \label{TR_PMTU}
\end{gather}
\textit{Let \(^DT_R\) be throughput with DMTU,}
\[\implies ^DT_R={Packet\ Size\over (^DT)}\]
\textit{Since, packet size = P,}
\begin{gather}
\implies ^DT_R={P\over (^DT)	} \label{TR_DMTU}	
\end{gather}

In the Equations \ref{TR_DMTU} and \ref{TR_PMTU} of DMTU and PMTUD throughput respectively we derive a graph by fixing a pre-defined value to all constant terms other than nodes. In other  words, the number of nodes varies in the route but keep other parameters constant. The value of constant terms will be highest/average limit of expected network simulator results or real time results. We carry this under different packet sizes so we take the iterations of packet size to see the rate of effect of increasing the packet size on the throughput by PMTUD and DMTU algorithms.

Since, on putting values of \(^DT\ and\ ^PT\) on Equation \ref{TR_DMTU} and \ref{TR_PMTU} respectively, we have:
\begin{gather}
^DT_R\; =\; \left( \frac{P}{n\left( ^PT_D+^{D-P}T_D \right)\; +\left( ^PT_D+^{D-P}T_D+aT_O \right)} \right)\\
^PT_R=\left(\frac{P}{\left( ^PT_D\left( n+1 \right)\right)\; +\; \left( ^PT_D+^PT'_D \right)\left( \sum_{i=1}^{a}{n_i} \right)+a\left( T_F \right)}\right)\;
\end{gather}

The pre-defined constant values are \(^PT_D=10^{-1},\;^PT'_D=^PT_D-\left( 15\cdot 10^{-3} \right),\; ^{D-P}T_D=10^{-2},\;  T_O=10^{-3},\; T_F=10^{-4}\)and let the packet size be constant i.e. \(P=\left( 1300 \right)\). Since, here we take varying value of percentage of nodes dropping the packet to get a clear view of how throughput acts with nodes drop packets.\\

\begin{figure}[!t]
  \centering
  \begin{subfigure}{0.4\columnwidth}
    \includegraphics[width=1\columnwidth]{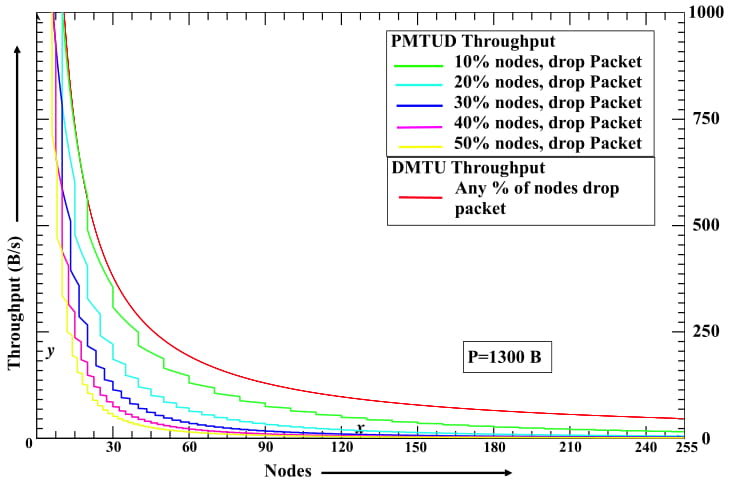}
    \caption{At Min(\(\sum_{i=1}^{\ a } [n_i]\))}
    \label{fig:putmin}
  \end{subfigure}
  \begin{subfigure}{0.4\columnwidth}
    \includegraphics[width=1\columnwidth]{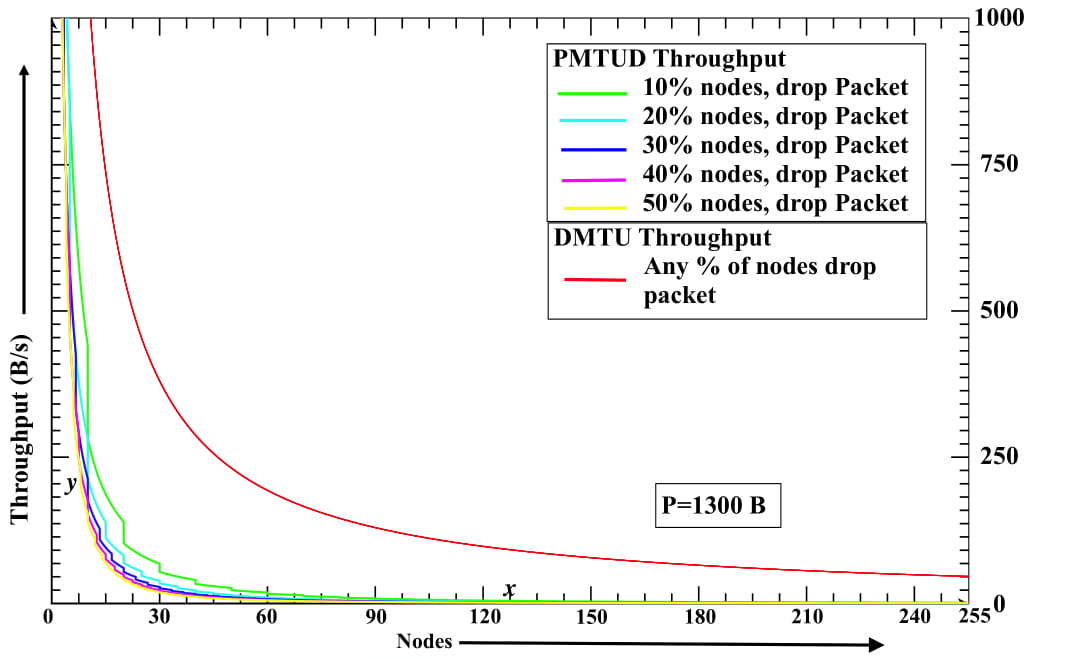}
    \caption{At Max(\(\sum_{i=1}^{\ a } [n_i]\))}
    \label{fig:putmax}
  \end{subfigure}
  \caption{Comparisons of throughput in DMTU and PMTUD algorithms.}
  \label{fig:coffee1}
\end{figure}

In Figure~\ref{fig:putmin} shows the throughput between PMTUD and DMTU at single packet at minimum value of \(\sum_{i=1}^{\ a } [n_i]\). For a single packet at different instant of percentage of nodes dropping packet the PMTUD throughput is lower  then DMTU while the DMTU throughput is smooth declining line with higher intercept from x-axis then PMTUD throughput. In other words the DMTU throughput is improved over the PMTUD throughput.

	The graph in the Figure~\ref{fig:putmax} shows the throughput between PMTUD and DMTU at single packet at maximum value of \(\sum_{i=1}^{\ a } [n_i]\). Which shows PMTUD throughput is getting lower for each increase in percentage of node drop packet than DMTU throughput. 
    Since, this graphical observation identifies that the throughput in DMTU algorithm is greater than in PMTUD algorithm for a given number of nodes \('n'\).
Again From Equation \ref{D<P} and \ref{D<P_Con}
we have,
\[^DT << ^PT\] 
\begin{gather}
\implies ^DT_R >\  ^PT_R	
\end{gather}

\textit{:. 	Throughput with DMTU \((^DT_R)\)  \(> \) Throughput with PMTUD \((^PT_R)\)}

\subsection{Graphical Analysis of Latency Enhancement}

Mathematically:
\[	Latency\ \alpha\  {1\over Throughput} \]

The latency is the inverse of throughput, therefore the throughput increases the latency decreases. 
The latency for the PMTUD algorithm is:
\begin{gather}
L_P	 =	K{1\over ^PT_R}
		=	K{^PT\over P} \label{Latency_PMTU}
\end{gather}
	
Where \(K\) is constant of proportionality. 
Now the latency for the DMTU algorithm will be:

\begin{gather}
        L_D	 =	K{1\over ^DT_R}
		=	K{^DT\over P} \label{Latency_DMTU}
\end{gather}

\begin{figure}[!t]
  \centering
  \begin{subfigure}{0.4\textwidth}
    \includegraphics[width=\hsize]{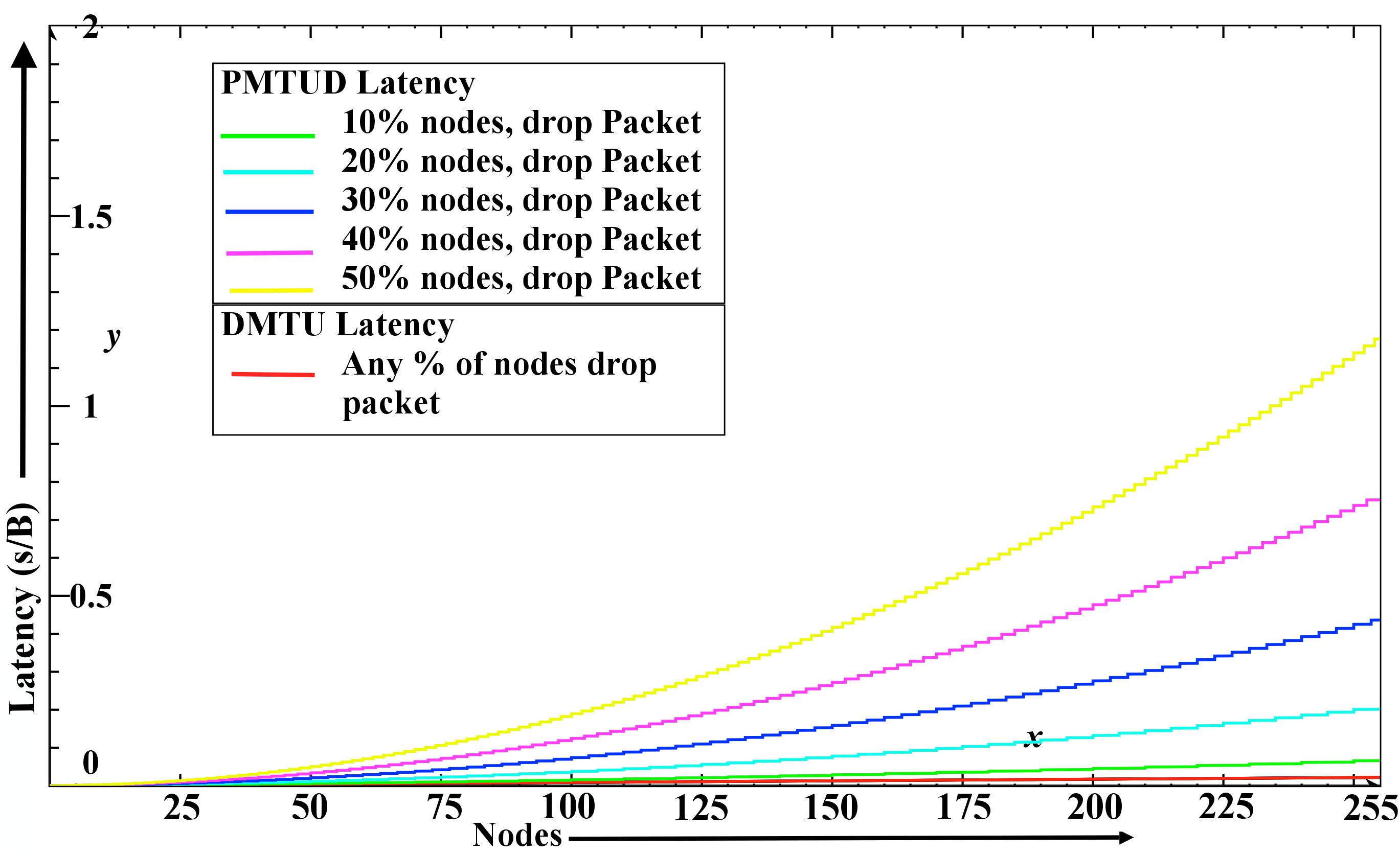}
    \caption{At Min(\(\sum_{i=1}^{\ a } [n_i]\))}
    \label{fig:laymin}
  \end{subfigure}
  \begin{subfigure}{0.4\textwidth}
    \includegraphics[width=\hsize]{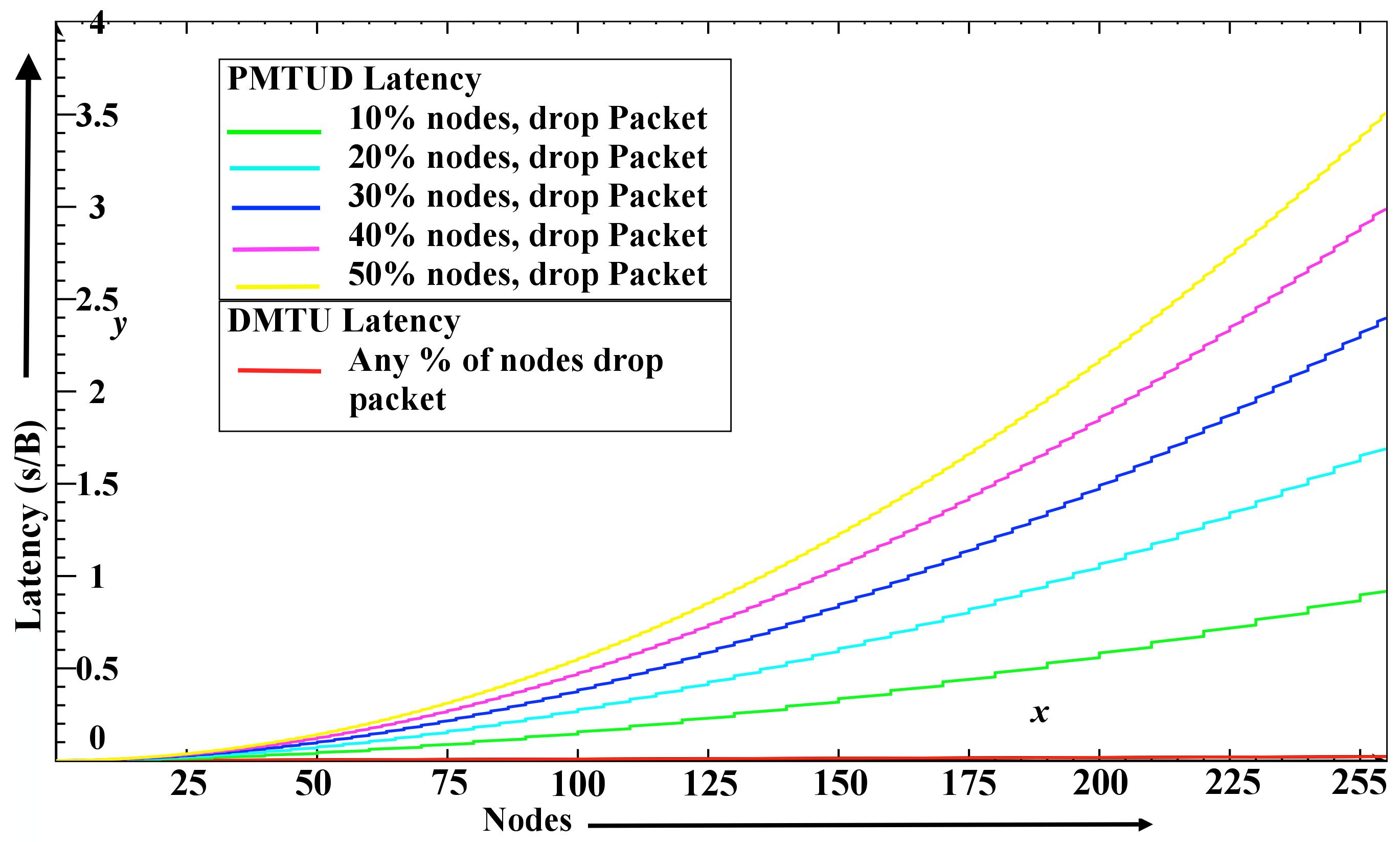}
    \caption{At Max(\(\sum_{i=1}^{\ a } [n_i]\))}
    \label{fig:laymax}
  \end{subfigure}
  \caption{Analysing Latency in DMTU and PMTUD algorithms.}
\end{figure}

let value of K=1 for accuracy and putting valued of \(^PT\ and\ ^DT\) in Equations \ref{Latency_PMTU} and \ref{Latency_DMTU} respectively which gives:
			\begin{gather}
			L_P=\left(\frac{\left( ^PT_D\left( n+1 \right)\right)\; +\; \left( ^PT_D+^PT'_D \right)\left( \sum_{i=1}^{a}{n_i} \right)+a\left( T_F \right)}{P}\; \label{laten_P}\right)\\
           L_D\; =\; \left(\frac{ n\left( ^PT_D+^{D-P}T_D \right)\; +\left( ^PT_D+^{D-P}T_D+aT_O \right)}{P}\; \label{laten_D}\right)
			\end{gather}
			
			we have taken the same pre-defined value for the above Equations \ref{laten_P} \& \ref{laten_D} i.e
			\(^PT_D=10^{-1},\;^PT'_D=^PT_D-\left( 15\cdot 10^{-3} \right),\; ^{D-P}T_D=10^{-2},\;  T_O=10^{-3},\; T_F=10^{-4}\). Like in analyses of throughput, hear we also taken a set of same order of packet size \(P=( 1300)\) with varying percentage of nodes dropping the packet.

In Figure~\ref{fig:laymin} shows the latency for the PMTUD algorithm and DMTU algorithm at minimum value of \(\sum_{i=1}^{\ a } [n_i]\).  The latency for PMTUD algorithm is increases with increase in percentage of nodes dropping packet while the latency for the DMTU algorithm is lower than the lowest Latency of PMTUD algorithm with a large Intercept of Y-axis.

Again in Figure~\ref{fig:laymax} shows the latency for the PMTUD algorithm and DMTU algorithm at maximum value of \(\sum_{i=1}^{\ a } [n_i]\). In maximum value of \(\sum_{i=1}^{\ a } [n_i]\) the Latency difference between the PMTUD algorithm and DMTU algorithm is larger than the minimum value of \(\sum_{i=1}^{\ a } [n_i]\).

Again by dividing Equation \ref{Latency_PMTU} with \ref{Latency_DMTU} we get:
\[	{L_P\over L_D}	=	{^PT\over ^DT}\]
Since, from Equation \ref{D<P} and \ref{D<P_Con} we have:
\begin{align}
&^DT <<\  ^PT \notag\\
\implies\qquad &{L_P\ \over L_D}\ >\ 1 \notag\\
\implies\qquad &L_P > L_D	
\end{align}

\textit{Hence the Latency with DMTU is Lower than Latency with PMTUD algorithm.}
\subsection{Routing Effectiveness}

\newtheorem{Lemma}{Lemma}
\begin{Lemma}
The ratio of Packet Delivery rate of PMTUD to Packet Delivery rate of DMTU is less than 1.\\
i.e.
\begin{align}
    {PDR_P \over PDR_d }< 1
\end{align}
\end{Lemma}

\begin{proof}
The number of packets delivered to the number of packets sent to reach to destination by source in same medium is given by Packet delivery rate (PDR) value:
\begin{align*}
  i.e. \qquad  &PDR = {Packet\ Received \over Packet\ Send}\\
   such\ that,\quad &( 0 \le \ (PDR) \ \le 1 )
    \end{align*}

If the PDR is near to 0, then it shows a less routing effectiveness of protocol and if PDR is near to 1, then it shows higher routing effectiveness of protocol.
We now calculate the PDR value for PMTUD and DMTU then compare them to see which has better routing effectiveness.
Let’s denote \(PDR_P\) for the PMTUD and \(PDR_D\) for DMTU.
Let x be number of packets to be send by source to destination with intermediate nodes n and links (n+1) both
for DMTU and PMTUD.
Let the packets drop due to network factors other than MTU be ‘C’ and the packets dropped due to MTU-size
would be \textit{‘l’}, then the Packets delivered will be \textit{x-(C+l)}

\begin{align}
\qquad PDR_P = {x-(C+l) \over x}\label{pdrp}
\end{align}

In DMTU the Packets drop due to factor of MTU or ‘Packet too big’ will be rescued. Therefore, $l=0$ , while the
packets drop due to other factors in network will be same as for PMTUD i.e.\textit{ ‘C’}.

\begin{align}
\label{pdrd}
\qquad PDR_D = {x-C \over x}
\end{align}

Dividing Equation \ref{pdrp} by \ref{pdrd}:
\begin{align}
\label{p/d}
{PDR_P\over PDR_D} = {x-(C+l) \over x -C}
\end{align}

In equation \ref{p/d} the numerator is less than denominator by value of \textit{ l} , therefore its ratio is less than 1.
\begin{align}
i.e \qquad &{PDR_P \over PDR_D} < 1\\
\implies \qquad &PDR_D > PDR_P
\end{align}

\textit{Hence, which shows that DMTU protocol has higher routing effectiveness than PMTUD.}
\end{proof}

%% file: issues.tex
\section{Issues of Implementation of Proposed Mechanism}
\label{sec:imp}
\subsection{Vendor restrictions}
The implementation of DMTU can be done by reprogramming the router's firmware with certain limits to change the MTU depending on the current packet being handled. These limits are kept to avoid the vulnerability that can be used maliciously. These limits were described in the mechanism. Its hard to implement on the router OS as its not been open for the educational or experimental use and due to restrict access to routers OS by the vendors we can't show the implementation in real world, that's why we are lacking the part of implementing the mechanism in this paper, which could only be done by the vendors side who have only the access to reprogram the router's firmware weather it could be physical or simulated router.

\subsection{Altering MTU - By rising MTU size}
Since the proposed method seems to violate the main purpose or definition of MTU in the network devices by altering it according to the incoming packet size. Then why shouldn't the MTU is removed, then to keep on overriding for each larger incoming packet? The proposed method has set a threshold limit of rising the MTU of the forwarding link and this threshold limit helps in keeping the main work of MTU unaffected. 

With the increase in link data rate and the new highly equipped modern equipment's the standard MTU 1500 bytes remain unchanged for decade which is now been in question by prior research carried out in support of using jumbo frames in IPv6 network by \cite{jumboipv6} shows the increase in network throughput by up to 117\% and rising the standard MTU 1500 bytes to 9000 bytes in \cite{largemtu} which found significantly increase in throughput and resistant to the packet drop due to reduce in the number of frame transmitted. In \cite{mtuEE} found that that the power consumption is proportional to the link load and with the larger MTUs allows a link to enter a low power mode, thereby improving energy efficiency in wire-line Ethernet.

\subsection{Pre-configuration for Implementation}
\subsection{Defining a Threshold Value}
Since the prior work in \cite{mtuEE}, \cite{largemtu}, \cite{jumboipv6} have advises to increase MTU to 9000 bytes and the current standard MTU is 1500 bytes. The threshold limit can be set between 1500 to 9000 bytes and keeping the standard 1500 MTU undisturbed. The threshold value can be used as the new prescribed standard MTU value for the Ethernet before implementing as standard MTU globally to help the high performance source and intermediate nodes and high link data with higher network performance as the older devices still can't work better in MTU size greater than 1500 bytes. The proposed method helps in sending high priority mini-jumbo and jumbo packets without disturbing standard MTU 1500 bytes.

\subsection{Halting restart - MTU change without restart }
In our proposed method the overriding of MTU will lead to connectivity failure as the change in MTU takes place after a automatic or manual restart of the network devices, to overcome this connectivity failure a pre-configuration is needed from vendor side by halting this feature of restarting the routers after a change in MTU and should make the change in MTU takes place without restart. This pre-configuration must be done before the implementation, for the smooth working of the proposed method.\\

These implementation issues are brought forward only in the view of designing the proposed method much more better by keeping the issues as the main factors in the knowledge of the researches who put their interest on making it more efficient and reliable.

%% file: conclusion.tex
\section{Conclusions \& Future Work}
\label{sec:conc}
This paper presents a scheme called DMTU that tries to reduce the packet drops
inside the network in order to increase the throughput and decrease the overall
packet latency. The algorithm proves to be robust and fast. This Method reduces usage of ICMP PTB messages in parallel Dynamic MTU scheme and nearly has null usage of ICMP PTB messages in Standalone Dynamic MTU. It is completely independent of the Source node  and any kind of Error messages until used in parallel with Path MTU Discovery Scheme as done by state-of-art algorithms in the same
domain.  Unlike PMTUD, the algorithm processes the packet as it arrives and
forwards in no time. This paper presents different versions of the DMTU
algorithm and shows how the minor optimizations help the algorithm to adapt to
different networks. The mathematical and graphical analyses of DMTU algorithm
shows its effectiveness as compared to the state-of-the-art PMTUD algorithm.

The DMTU algorithm can be further enhanced in terms of the overhead time it takes to process the packet and overrides the MTU of a node which will have a great effect in the complete implementation and working of the algorithm and hence thereby decreases the time delay further. Our main motive of designing the DMTU algorithm is to get rid of the Packet truncation and congestion in IPv6 protocol networking which effects the Quality of Service of the network.

The further work in finding the time complexity and space complexity of the program that takes to run algorithm program on end-devices and intermediate node.


%% file: main.bbl
\begin{thebibliography}{10}

\bibitem{rfc4632}
V~Fuller and T~Li.
\newblock Classless inter-domain routing (cidr): The internet address
  assignment and aggregation plan.
\newblock RFC 4632, IETF, Internet Requests for Comments, 8 2006.

\bibitem{rfc8200}
Steve Deering and Robert~M. Hinden.
\newblock Internet protocol, version 6 (ipv6) specification.
\newblock RFC 8200, IETF, Internet Request for Comments, 7 2017.

\bibitem{rfc4891}
Richard Graveman, Pekka Savola, Mohan Parthasarathy, and Hannes Tschofenig.
\newblock Using ipsec to secure ipv6-in-ipv4 tunnels.
\newblock RFC 4891, IETF, Internet Request for Comments, 5 2007.

\bibitem{tunnel}
P.~{Wu}, Y.~{Cui}, J.~{Wu}, J.~{Liu}, and C.~{Metz}.
\newblock Transition from ipv4 to ipv6: A state-of-the-art survey.
\newblock {\em IEEE Communications Surveys Tutorials}, 15(3):1407--1424, 2013.

\bibitem{rfc4291}
Robert~M. Hinden and Steve Deering.
\newblock Ip version 6 addressing architecture.
\newblock RFC 4291, IETF, Internet Requests for Comments, 2 2006.

\bibitem{IEEE}
IEEE.
\newblock Ieee standards for local area networks: Carrier sense multiple access
  with collision detection (csma/cd) access method and physical layer
  specifications.
\newblock {\em ANSI/IEEE Std 802.3-1985}, 1985.

\bibitem{conjestion}
Noor Mast and Thomas~J. Owens.
\newblock A survey of performance enhancement of transmission control protocol
  (tcp) in wireless ad hoc networks.
\newblock {\em EURASIP Journal on Wireless Communications and Networking},
  2011(1):96, 2011.

\bibitem{errorR}
R.~{Khalili} and K.~{Salamatian}.
\newblock A new analytic approach to evaluation of packet error rate in
  wireless networks.
\newblock {\em IEEE Proceedings of the 3rd Annual Communication Networks and
  Services Research Conference (CNSR'05)}, pages 333--338, 2005.

\bibitem{collision}
Buddha Singh and D.~K. Lobiyal.
\newblock A mac-layer retransmission technique for collided packets in wireless
  sensor network.
\newblock {\em Wireless Personal Communications}, 72(4):2499--2518, 2013.

\bibitem{rfc8201}
Jack McCann, Stephen~Edward Deering, Jeffrey~C. Mogul, and Robert~M. Hinden.
\newblock Path mtu discovery for ip version 6.
\newblock RFC 8201, IETF, Internet Request for Comments, 7 2017.

\bibitem{frag}
Christopher~A. Kent and Jeffrey~C. Mogul.
\newblock Fragmentation considered harmful.
\newblock {\em SIGCOMM Comput. Commun. Rev.}, 25(1):75–87, January 1995.

\bibitem{BB12}
Maikel~De Boer, Jeffrey Bosma, B~Overeinder, and Willem Toorop.
\newblock Discovering path mtu black holes on the internet using ripe atlas.
\newblock Master's thesis, University of Amsterdam, The Netherlands, 7 2012.

\bibitem{rfc1191}
Jeffrey~C. Mogul and Stephen~Edward Deering.
\newblock Path mtu discovery.
\newblock RFC 1191, IETF, Internet Requests for Comments, 11 1990.

\bibitem{PMTU-ISIS}
Vijay~Kumar Vasantha.
\newblock Isis: Path mtu calculation in isis.
\newblock Internet Draft draft-kumar-isis-path-mtu-00.txt, IETF, 8 2008.

\bibitem{PMTU-ISISE}
Z~Hu, Y~Zhu, Z~Li, and L~Dai.
\newblock Is-is extensions for path mtu draft-hu-lsr-isis-path-mtu-00.
\newblock Technical report, IETF Internet-Draft, 6 2018.

\bibitem{rfc4821}
M.~Mathis and J.~Heffner.
\newblock Packetization layer path mtu discovery.
\newblock RFC 4821, IETF, Internet Requests for Comments, 2007.

\bibitem{PSC}
Matt Mathis.
\newblock Raising the internet mtu.
\newblock {\em Pittsburgh supercomputing center}, 04 2007.

\bibitem{rfc8899}
G.~Fairhurst, Tom Jones, M.~T{\"u}xen, Irene R{\"u}ngeler, and T.~V{\"o}lker.
\newblock Packetization layer path mtu discovery for datagram transports.
\newblock RFC 8899, IETF, Internet Requests for Comments, 09 2020.

\bibitem{dos}
Tasnuva Mahjabin, Yang Xiao, Guang Sun, and Wangdong Jiang.
\newblock A survey of distributed denial-of-service attack, prevention, and
  mitigation techniques.
\newblock {\em International Journal of Distributed Sensor Networks},
  13(12):1550147717741463, 2017.

\bibitem{dosaic}
Khaled Elleithy, Drazen Blagovic, Wang Cheng, and Paul Sideleau.
\newblock Denial of service attack techniques: Analysis, implementation and
  comparison.
\newblock {\em Journal of Systemics, Cybernetics and Informatics}, 3:66--71, 01
  2006.

\bibitem{PMTUDUNI}
Matthew Luckie, Kenjiro Cho, and Bill Owens.
\newblock Inferring and debugging path mtu discovery failures.
\newblock 2005.

\bibitem{rfc1435}
S~Knowles.
\newblock Iesg advice from experience with path mtu discovery.
\newblock RFC 1435, IETF, Internet Requests for Comments, 3 1993.

\bibitem{rfc2923}
K~Lahey.
\newblock Tcp problems with path mtu discovery.
\newblock RFC 2923, IETF, Internet Requests for Comments, 9 2000.

\bibitem{rfc4459}
P~Savola.
\newblock Mtu and fragmentation issues with in-the-network tunneling.
\newblock RFC 4459, IETF, Internet Requests for Comments, 4 2006.

\bibitem{rfc791}
Jon Postel.
\newblock Internet protocol.
\newblock RFC 791, IETF, Internet Request for Comments, 9 1981.

\bibitem{rfc7719}
P~Hoffman, A~Sullivan, and K~Fujiwara.
\newblock Dns terminology.
\newblock RFC 7719, IETF, Internet Request for Comments, 12 2015.

\bibitem{dns}
K.~{Rikitake}, H.~{Nogawa}, T.~{Tanaka}, K.~{Nakao}, and S.~{Shimojo}.
\newblock Dns transport size issues in ipv6 environment.
\newblock {\em 2004 International Symposium on Applications and the Internet
  Workshops. 2004 Workshops, IEEE.}, pages 141--145, 2004.

\bibitem{rfc1981}
J~McCann, S~Deering, and J~Mogul.
\newblock Path mtu discovery for ip version 6.
\newblock RFC 1981, IETF, Internet Requests for Comments, 8 1996.

\bibitem{ishfaq}
Ishfaq Hussain and Janibul Bashir.
\newblock Measuring time delay in path mtu discovery in transmitting a packet
  in ipv4 and ipv6 network.
\newblock 2020.

\bibitem{jumboipv6}
{Supriyanto}, R.~{Sofhan}, R.~{Fahrizal}, and A.~{Osman}.
\newblock Performance evaluation of ipv6 jumbogram packets transmission using
  jumbo frames.
\newblock {\em 2017 4th International Conference on Electrical Engineering,
  Computer Science and Informatics (EECSI) [proceedings]}, pages 1--5, 2017.

\bibitem{largemtu}
D.~{Murray}, T.~{Koziniec}, K.~{Lee}, and M.~{Dixon}.
\newblock Large mtus and internet performance.
\newblock pages 82--87, 2012.

\bibitem{mtuEE}
P.~Reviriego, A.~Sanchez-Macian, J.A. Maestro, and Chris~J. Bleakley.
\newblock Increasing the mtu size for energy efficiency in ethernet.
\newblock {\em IET Irish Signals and Systems Conference (ISSC 2010)
  [proceedings]}, 06 2010.

\end{thebibliography}
